\documentclass[preprint,review,10pt]{elsarticle}
\usepackage{color,soul}
\usepackage{hyperref}
\usepackage[english]{babel}
\usepackage{ragged2e}
\usepackage{blindtext}
\usepackage{caption}
\usepackage{subcaption}
\usepackage{multirow}
\usepackage{graphicx}
\usepackage[symbol]{footmisc}
\usepackage{url}

\usepackage{amssymb}

\journal{Renewable Energy}

\begin{document}

\begin{frontmatter}



\title{Multi-model assessment of heat decarbonisation options in the UK using electricity and hydrogen}

\author[inst1]{Marko Aunedi\corref{cor1}}
\cortext[cor1]{ Marko Aunedi, Maria Yliruka, and Shahab Dehghan contributed equally to this work.}
\affiliation[inst1]{organization={Imperial College London},
            addressline={South Kensington Campus}, 
            city={London},
            postcode={SW7 2AZ}, 
            country={UK}}

\author[inst1]{Maria Yliruka\corref{cor1}}
\author[inst1]{Shahab Dehghan\corref{cor1}}
\author[inst1,inst2]{Antonio Marco Pantaleo}
\author[inst1]{Nilay Shah}
\author[inst1]{Goran Strbac}

\affiliation[inst2]{organization={University of Bari Aldo Moro},
            addressline={Piazza Umberto I 1}, 
            city={Bari},
            postcode={70121}, 
            country={Italy}}

\begin{abstract}
Delivering low-carbon heat will require the substitution of natural gas with low-carbon alternatives such as electricity and hydrogen. The objective of this paper is to develop a method to soft-link two advanced, investment-optimising energy system models, RTN (Resource-Technology Network) and WeSIM (Whole-electricity System Investment Model), in order to assess cost-efficient heat decarbonisation pathways for the UK while utilising the respective strengths of the two models. The linking procedure included passing on hourly electricity prices from WeSIM as input to RTN, and returning capacities and locations of hydrogen generation and shares of electricity and hydrogen in heat supply from RTN to WeSIM. The outputs demonstrate that soft-linking can improve the quality of the solution, while providing useful insights into the cost-efficient pathways for zero-carbon heating. Quantitative results point to the cost-effectiveness of using a mix of electricity and hydrogen technologies for delivering zero-carbon heat, also demonstrating a high level of interaction between electricity and hydrogen infrastructure in a zero-carbon system. Hydrogen from gas reforming with carbon capture and storage can play a significant role in the medium term, while remaining a cost-efficient option for supplying peak heat demand in the longer term, with the bulk of heat demand being supplied by electric heat pumps.
\end{abstract}

\begin{graphicalabstract}
\includegraphics[width=\linewidth]{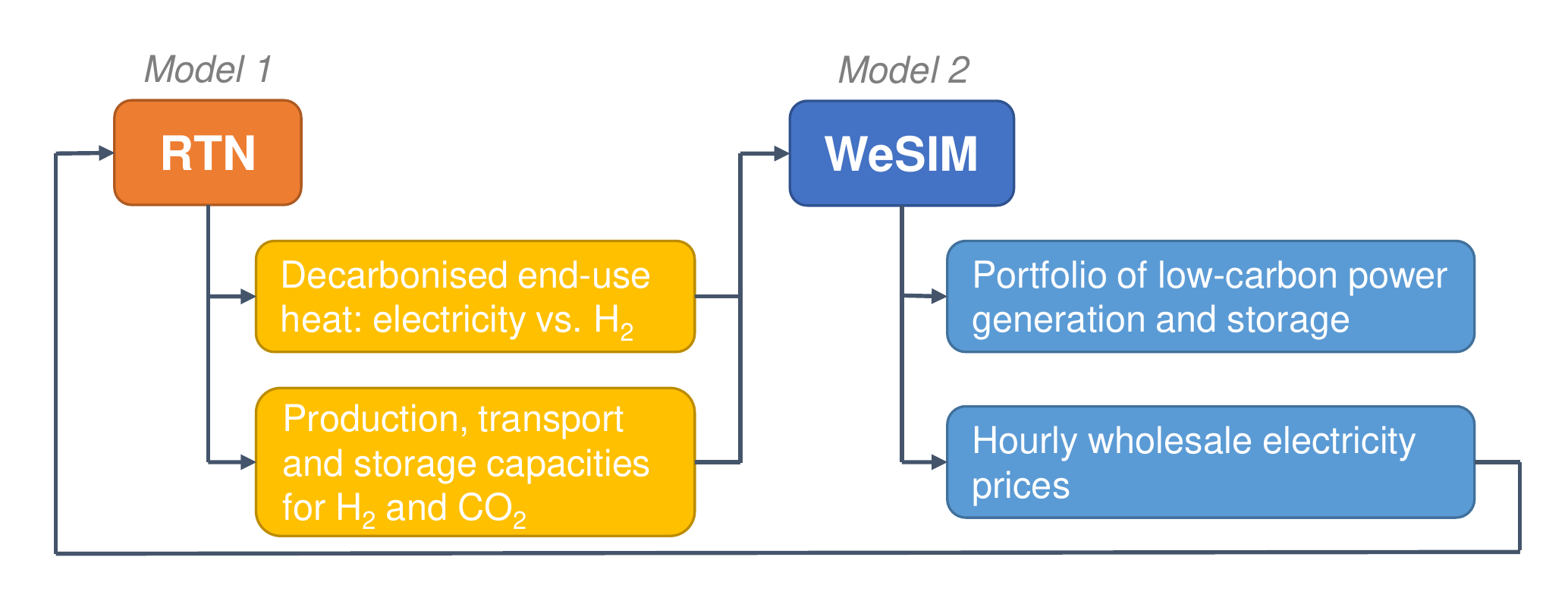}
\end{graphicalabstract}

\begin{highlights}
\item A multi-model approach is proposed to link two established energy system models. 
\item Iterating between the two models focused on low-carbon heat and electricity sectors. 
\item Various technologies are considered for electricity and heat decarbonisation.
\item Contributions of hydrogen and electricity to heat supply are assessed until 2050.
\item Evolution of the cost-effective portfolio of hydrogen sources is characterised.
\end{highlights}


\end{frontmatter}



\section*{List of abbreviations}
\label{sec:ListOfAbbreviations}

\begin{tabular}{l l}
    ASHP & Air Source Heat Pump \\
    ATR & Autothermal Reformer \\
    CCGT & Combined-Cycle Gas Turbine \\
    CCS & Carbon Capture and Storage \\
    COP & Coefficient of Performance \\
    DHN & District Heat Network \\
    DSR & Demand-Side Response \\
    FES & Future Energy Scenarios \\
    GB & Great Britain \\
    GHR & Gas-Heated Reformer \\
    HP & Heat Pump \\
    LOLE & Loss of Load Expectation \\
    LP & Linear Programming \\
    MILP & Mixed-Integer Linear Programming \\
    NPV & Net Present Value \\
    OCGT & Open-Cycle Gas Turbine \\
    PV & Photovoltaics \\
    PEM & Proton-Exchange Membrane \\
    RES & Renewable Energy Sources \\
    RTN & Resource-Technology Network \\
    SMR & Steam Methane Reformer \\
    SOE & Solid Oxide Electrolyser \\
    UK & United Kingdom \\
    WeSIM & Whole-electricity System Investment Model
\end{tabular}

\section{Introduction}
\label{sec:Introduction}

This section outlines the motivation and background for the analysis presented in the paper, presenting the review of relevant literature. It also highlights the key contributions of the paper and lays out its structure.

\subsection{Motivation and Background}
\label{subsec:Motivation and Background}
A wide-ranging transformation is required in energy systems around the world over the next three decades to meet the ambitious targets set for decarbonisation \cite{stanvcin2020review,papaefthymiou2016towards}. In 2019, the United Kingdom became the first major economy in the world to adopt a law aimed at bringing the greenhouse gas emissions to net-zero by 2050 \cite{BEIS2019}. The UK’s heat sector, predominantly supplied by natural gas at present, accounts for a third of the total carbon emissions \cite{BEIS2019EM,HOP2016}. 

Likely pathways for heat decarbonisation as the most challenging energy sector to meet the net-zero target involve substituting fossil fuels with low-carbon energy vectors, electricity and hydrogen, produced from renewable or other forms of zero-carbon energy \cite{lowes2020disruptive}. Renewable energy systems have recently experienced significant progress with respect to technological development, resource assessment, cost reduction and system design \cite{OSTERGAARD20202430}. Specific recent advances in RES reported in the literature include hybridisation and cross-sector integration \cite{OSTERGAARD20201554} as well as building-integrated systems and novel resource assessment approaches \cite{OSTERGAARD2021877}.

The UK Government's Net Zero Strategy \cite{HMGov2021NetZero} set the ambition that by 2035 all new heating appliances installed in homes and workplaces will be low-carbon technologies, like electric heat pumps or hydrogen boilers. At the same time, the Strategy also envisages a full decarbonisation of electricity supply by 2035 and 5~GW of low-carbon hydrogen production capacity by 2030.

At present, only about 1\% of the UK’s 28 million homes are heated by heat pumps (HPs). The UK Government’s Ten Point Plan for Green Industrial Revolution \cite{HMGov2020TenPointPlan} set the objective of installing 600,000 HPs per year in the period 2021-2028. Long-term projections for the UK energy system generally envisage a very high future penetration of HP systems; for instance, one of National Grid’s Future Energy Scenarios \cite{NGFES2020} (Consumer Transformation) projects up to 26.2 million HP systems (including hybrids) in the UK by 2050, which is equivalent to almost every household in the UK having a HP system installed.

Achieving the UK’s long-term climate targets cost-effectively requires a coordinated approach to decarbonise both heat and electricity supply \cite{sansom2014decarbonising,bistline2021role}. Previous research has demonstrated that heating and electricity systems can benefit significantly from mutual synergies on their pathways towards decarbonisation, by unlocking opportunities for cross-vector flexibility to support the integration of low-carbon generation technologies and to significantly reduce the cost of decarbonisation \cite{Zhang2018,aunedi2020modelling,jimenez2020coupling,pavivcevic2019comparison,pavivcevic2020potential,dominkovic2016zero,CCC2018,ramsebner2021sector}. In other words, integrated and coordinated design of heat and electricity supply systems is likely to result in a lower low-carbon transition cost compared to a ``silo'' approach optimising each supply system in a decoupled fashion.

It has been shown that a cost-efficient decarbonisation of energy supply through large volumes of variable renewables requires a variety of technologies to provide flexibility in the context of grid support, balancing, and adequacy or security of supply \cite{Strbac2020PiE}, aimed at continuous production and consumption balance within the whole energy system. These technologies include various forms of energy storage, demand-side response, expansion of interconnection capacity and more flexible generation technologies, as well as a number of cross-vector flexibility or sector coupling options, such as hydrogen production and conversion technologies \cite{pudjianto2013whole,gjorgievski2021potential,liu2020review}. 

Many previous studies have highlighted the critical role of flexibility, particularly in low-inertia power systems with very high penetrations of renewable energy resources with the objective to achieve net zero or even net negative carbon emission targets \cite{gjorgievski2021potential,davis2018net,IDLES2021}.

The cost-effectiveness of hydrogen as a flexible intermediate energy vector in a highly renewable energy system has been previously demonstrated in \cite{fu2020integration}, which estimated the potential benefits of deploying flexible hydrogen infrastructure in supplying both heat and electricity sectors at several billion pounds of savings in annual energy system cost. The transition to a low-carbon energy system would rely on various production technologies for green and blue hydrogen, including electrolysers, biomass gasification and reformers, e.g., Steam Methane Reforming (SMR) and Autothermal Reforming (ATR), coupled with Carbon Capture and Storage (CCS) and negative emission technologies \cite{kovavc2021hydrogen}. 

At the European level there are also expectations towards a significant increase of hydrogen use in transport, industry and the building (heating) sector, where hydrogen can gradually substitute the current use of natural gas \cite{Trinomics2020}. In the global context, hydrogen is recognised as a potential solution for sectors that are hard to electrify, providing a link between low-carbon electricity generation and the hydrogen demand sectors, although a number of barriers still exist for its widespread uptake \cite{IRENA2020}. Although challenges around the cost and performance of hydrogen technologies remain, they could become competitive in the medium term, which justifies the renewed interest and policy support for these technologies around the world \cite{staffell2019role}.

Currently, hydrogen is mainly produced using reforming technologies \cite{stanvcin2020review}. However, biomass gasification and water electrolysis represent viable alternatives for producing green hydrogen in an energy system with a high share of renewable resources, and their costs have been declining \cite{liu2020review,kovavc2021hydrogen}. Recent cost estimates foresee a further reduction in the cost of hydrogen production technologies \cite{BEIS2021H2}.

Several recent publications have focused on integrating and utilising hydrogen in the context of the whole energy system. In \cite{damen2007comparison}, a hydrogen supply chain including production, storage, and distribution technologies is developed for the UK transport demand. In \cite{almansoori2006design}, another hydrogen supply chain is designed by optimising infrastructural and operational costs for the transport sector. In \cite{fu2021integration}, a large-scale model is presented for optimising the interactions between electricity, hydrogen, and transport sectors and identify cost-efficient decarbonisation scenarios for the whole energy system in the UK. 

In \cite{colbertaldo2018modelling}, a multi-node model is introduced to design the low-carbon integrated electricity and transport network in Italy while considering power-to-gas technologies that utilise excess electricity in zero-carbon transport. Another multi-node model is developed in \cite{welder2018spatio} to investigate potential cost-effective low-carbon pathways for the future integrated energy system in Germany focusing on the integration of power-to-gas technologies, where the time-dependent characteristics of the integrated energy system are modelled by representative periods obtained by clustering techniques. 

Although characterising the time-dependent nature of integrated energy systems through a limited number of representative periods simplifies computational issues, it may result in significant errors in capturing the requirements for flexibility in the energy system due to oversimplified temporal variations of both energy demand and intermittent renewable energy generation.

Modelling and optimising low-carbon energy systems with high penetration of intermittent renewable energy resources (particularly wind and solar) need both a technology-rich representation of energy generation, transport, storage and end-user consumption under different techno-economic constraints, as well as a sufficiently accurate representation of spatial and temporal variations in energy supply and energy demand. In practice, simulating and analysing multi-vector energy systems with a high spatio-temporal resolution is computationally cumbersome \cite{collins2017integrating}. 

To address these challenges, several approaches have been reported in the literature focusing on soft-linking multiple long-term and short-term models for optimising energy system investment and operation models in order to obtain a tractable energy planning problem offering a higher level of accuracy compared to separate utilisation of each model. Examples include soft-linking the JRC-EU-TIMES and the Dispa-SET models in \cite{pavivcevic2020potential}, the JRC-EU-TIMES model and a behavioural model for transportation in \cite{blanco2019soft}, and TIMES and EMEC models in \cite{krook2017challenges}. 

Multi-model approaches provide new insights that would not be possible using a single model approach; for instance, the authors of \cite{collins2017adding} reported that complementing the PRIMES energy systems model with an EU-28 electricity dispatch model suggested that PRIMES overestimated the variable renewable generation and underestimated curtailments and grid congestions.

Previous studies on model coupling have mostly focused on ``unidirectional'' soft-linking of long-term and short-term energy models. However, to the best of the authors' knowledge, there is little research available in the literature with bidirectional result exchange between multiple investment optimisation models as a system of systems to decarbonise integrated heat and electricity sectors with sufficiently high spatial and temporal resolution under technology-rich representation. 

Therefore, to enhance accuracy and ensure tractability, a multi-model approach is proposed in this paper to study the decarbonision of integrated heat and electricity sectors in the UK through bidirectionally linking the Resource-Technology Network (RTN) model \cite{pantelides1994unified}, characterised by high spatial resolution and technological detail for hydrogen supply network, and the Whole-electricity System Investment Model (WeSIM) that represents the power system with high temporal resolution and a high level of technical detail \cite{pudjianto2013whole}.

RTN model is a spatio-temporal Mixed-Integer Linear Programming (MILP) model that has been used previously to assess the potential of hydrogen in the decarbonisation of the UK’s heating sector. For instance, a model based on wind power is presented in \cite{samsatli2019role} to supply domestic heating demand through either electric heaters or hydrogen production via water electrolysis. The authors reported that the selection of hydrogen-based technologies was sensitive to the availability of large-scale hydrogen storage. 

A similar observation was made in \cite{sunny2020needed}, which further concluded that cost-optimal regions for deploying hydrogen infrastructure for heating tend to be characterised by higher heating demand and located in proximity to hydrogen and $\mathrm{CO}_2$ storage sites. The same study also concluded that the use of large-scale hydrogen storage and deep geological reservoirs for $\mathrm{CO}_2$ would allow for a cost-effective transformation of the incumbent natural gas-based heat supply system to hydrogen.

RTN allows for developing long-term energy system strategies through multi-period analysis. In the context of decarbonisation, it enables the identification of cost-efficient pathways to transform the existing energy system and reach the goal of net-zero emissions. Different options have been considered to transition from current fossil fuel-based systems, including the electrification of heat and transport sectors. 

In the domestic transport sector, the potential of new onshore wind farms to supply zero-carbon transport demand has been explored for electric \cite{samsatli2016whole} and for hydrogen-based vehicles \cite{samsatli2016optimal}. On a district level, carbon-constrained cost-optimisation models based on the RTN framework have explored decentralised systems for domestic and commercial heat and electricity supply \cite{keirstead2012impact,keirstead2012capturing}.

The other model that is used for assessing the heat decarbonisation options in this paper is the Whole-electricity System Investment Model (WeSIM), presented in \cite{pudjianto2013whole}. WeSIM has been used extensively to study the challenges of integrating large volumes of low-carbon generation in future electricity systems, including the role and value of flexible technologies such as smart EV charging \cite{aunedi2013efficient}, Vehicle-to-Grid \cite{aunedi2020whole}, battery storage \cite{strbac2017opportunities}, pumped-hydro storage \cite{teng2018assessment} and liquid-air and pumped-heat energy storage \cite{georgiou2020value}. 

The objective of this paper is to take advantage of the two models’ strengths to assess heat decarbonisation pathways for the UK, i.e. to utilise the high geographical detail and the range of hydrogen production and end-use heating technological options considered in RTN and combine them with a high temporal resolution and detailed representation of the electricity system enabled by WeSIM. 

To the best of authors’ knowledge, there is no multi-model approach in the literature based on interactions between multiple models for heat and electricity sectors with high spatial and temporal resolutions, as presented in this paper, that determine both the design as well as operation of a low-carbon energy system with hydrogen-based technologies.

\subsection{Contributions}
\label{subsec:Contributions}
In light of the above, the key proposed contributions of the paper are the following:
\begin{enumerate}
    \item Develop a multi-modelling approach to study pathways for cost-efficient decarbonisation of heat supply using electricity and hydrogen;
    \item Propose a methodological framework for soft-linking two established energy system models with different geographical scopes, temporal resolutions and technology coverage;
    \item Determine the cost-effective technology mix for delivering zero-carbon heat through a combination of electrification and hydrogen-based heating.
\end{enumerate}
\subsection{Paper Structure}
\label{subsec:PaperStructure}
The remainder of the paper is organised as follows. In Section \ref{sec:Method}, the proposed method for linking the RTN and WeSIM models is discussed in detail. Section \ref{sec:ResultsDiscussion} presents and discusses different case studies for decarbonising the integrated heat and electricity sectors in the UK by 2050. Finally, Section \ref{sec:Conclusion} summarises the main conclusions of the paper.

\section{Method}
\label{sec:Method}
This section presents the multi-model assessment approach used in this paper and outlines the key features of RTN and WeSIM model. It also summarises the key assumptions and scenarios used in the quantitative studies.

\subsection{Rationale for multi-model assessment}
\label{subsec:Rationale}
The multi-vector models proposed in literature generally do not simultaneously deal with issues such as technological richness, computational complexity associated with fine temporal scales, coordination of district and national objectives, uncertainty, and multiple agent perspectives including the interactions between national and local energy systems \cite{samsatli2015bvcm,pfenninger2014energy,welder2018spatio}. 

In order to overcome the limits of the existing energy system models, the IDLES research programme \cite{IDLESwebsite}, which also includes the analysis presented in this paper, aims to propose a framework for creating a system-of-systems model that is computationally tractable when dealing with multi-physics models across energy carriers, with high temporal resolution (to capture temporal features of technology and system operation) and with high spatial resolution (for accurate reflection of national and local impact of various types of energy infrastructure). The ambition of this framework is to achieve technology richness sufficient to inform technology development goals, establishing the value that different technologies can provide to the energy system and how this relates to the infrastructure planning, technologies design and operational strategies.

In this context, the key aim of this paper is to combine the high spatial resolution, multi-vector and sector coupling features of RTN with the high temporal resolution and detailed representation of the power system of WeSIM. The multi-model approach allows for an enhanced representation of the complex interactions associated with energy system design and operation at both local and national level, quantifying with higher accuracy the aspects such as the benefits of flexibility, coupling between electricity, heat and hydrogen, energy storage and ancillary services. 

Moreover, the integrated approach that combines the capabilities of RTN and WeSIM models could better identify and quantify the system-wide benefits of technologies and infrastructures that cross the boundaries between electricity, heat and hydrogen supply systems.

The areas of investigation proposed in this application include heat decarbonisation technologies via electric and/or hydrogen systems and the consequent implications in terms of energy system flexibility and storage needs to enable a high penetration of intermittent renewable energy sources for net-zero energy scenarios by 2050. However, the proposed multi-model approach could be suitable to appreciate the system value of other zero-carbon heating technologies as well as inter-seasonal heat storage.

\subsection{Key features of RTN}
\label{subsec:RTNfeatures}
In this work, the RTN model presented in \cite{sunny2020needed} is extended and adapted for the integration with WeSIM. It simultaneously optimises the design and operation of hydrogen and $\mathrm{CO}_2$ infrastructures to supply domestic and commercial heating demand. For this work, the pressure levels in hydrogen pipelines are however omitted. 

\begin{figure}[t!] 
	\centering
    \includegraphics[width=\linewidth]{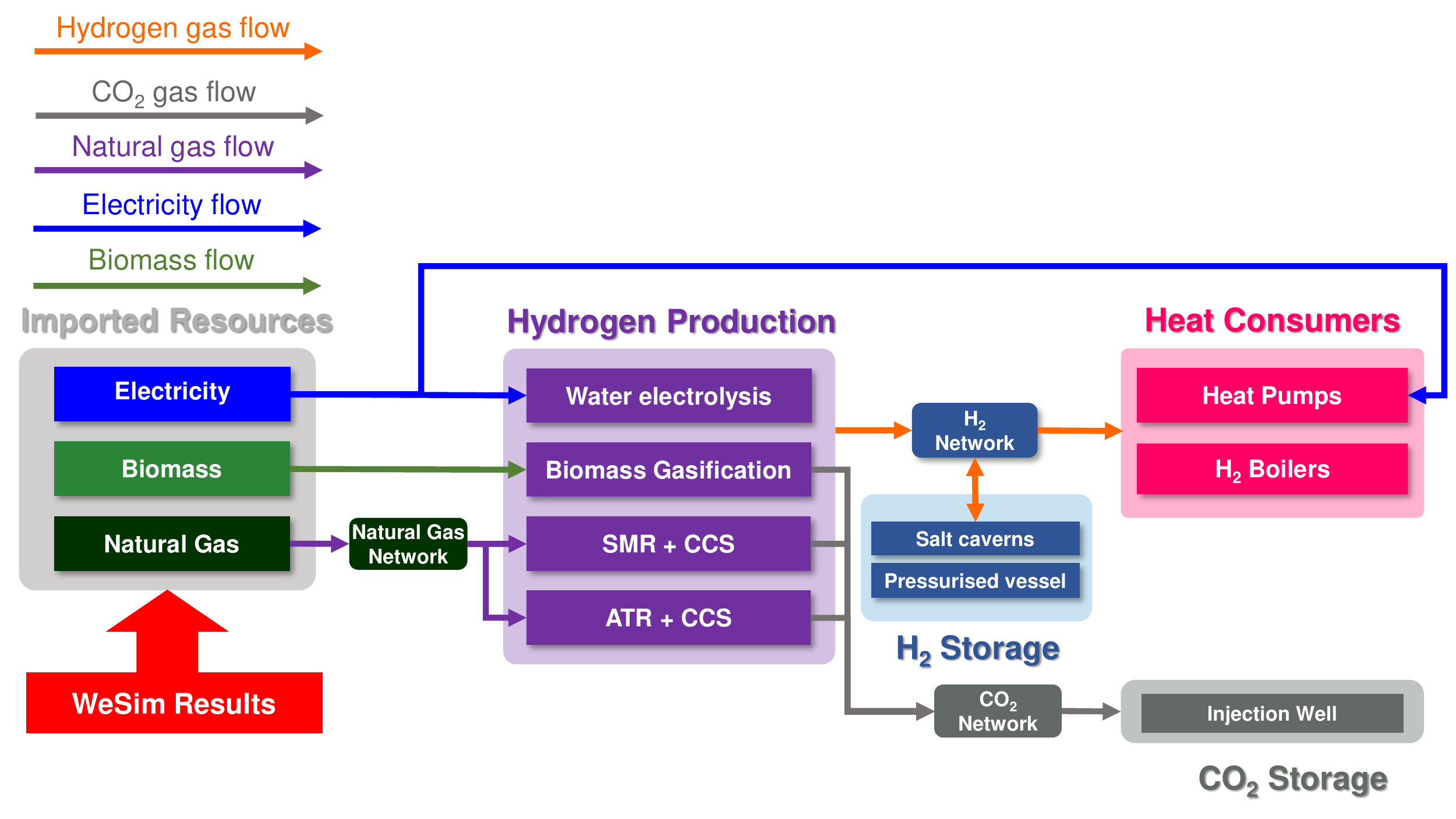}
	\caption{{\color{black}High-level diagram of RTN model}}
	\label{fig: RTN model}
\end{figure}

In the general RTN formulation, any material or energy stream is a resource that can be consumed, produced, or stored using different technologies. Figure~\ref{fig: RTN model}  illustrates the main components of the model: imported resources/fuels, various hydrogen production technologies, hydrogen and $\mathrm{CO}_2$ storage technologies, and end-use heating technologies. 

While the generation of hydrogen and $\mathrm{CO}_2$ is explicitly modelled, electricity, biomass and natural gas can be imported. Electricity is the only imported resource that can be used directly to provide heat via heat pumps, without explicitly considering the need for dedicated infrastructure. All three resources can also be used to produce hydrogen via water electrolysis, biomass gasification or reforming of natural gas coupled with CCS (SMR or ATR). 

The captured $\mathrm{CO}_2$ is transported and stored at offshore sites. If the produced hydrogen is not consumed locally, it can be transported via pipelines and stored in salt caverns for intra-day and inter-seasonal storage. Pressurised vessels are considered as additional hydrogen storage. Each technology is characterised by its capital cost, conversion efficiency, lifetime and operational costs. For the cost of the resources, the model distinguishes between wholesale and retail prices for electricity, natural gas and hydrogen.

Similar to the widely used UK TIMES model \cite{UKTIMESwebsite}, the RTN model uses 16 time slices to represent temporal variations in demand. These include four representative daily periods across three seasons (winter, autumn/spring and summer) plus the explicit consideration of the peak winter heat demand day to ensure correct sizing of the supply infrastructure capacity. The four daily time slices correspond to daytime, evening peak, late evening, and night periods. 

The full time horizon of 2030-2060 considered in the optimisation is divided into three major investment periods, each consisting of 10 years. The annual heat demand and electricity price profiles are aggregated based on hourly profiles and therefore vary across different time slices. While the biomass price is assumed to remain constant throughout the planning time horizon, the price of natural gas is assumed to vary seasonally. 

For the supply chain optimisation, the geographical area of Great Britain is divided into 51 equally sized cells using the open-source Geographic Information System QGIS \cite{QGIS}. Each cell is characterised by the heat demand profile of the domestic and commercial sector, its distance to neighbouring cells, available geological hydrogen storage capacity and existing natural gas transmission lines. The offshore sites for  $\mathrm{CO}_2$ storage are considered by specifying three additional grid cells, one located in the East Irish Sea and two in the North Sea.

The objective function in RTN is to minimise the total Net Present Value (NPV) of system cost over the entire time horizon, comprising the investment cost of technologies chosen for installation and the operating costs that are calculated based on the cost of imported resources and operating parameters of the technologies. The full RTN formulation is presented in \cite{sunny2020needed}.

\begin{figure}[t!] 
	\centering
    \includegraphics[width=\linewidth]{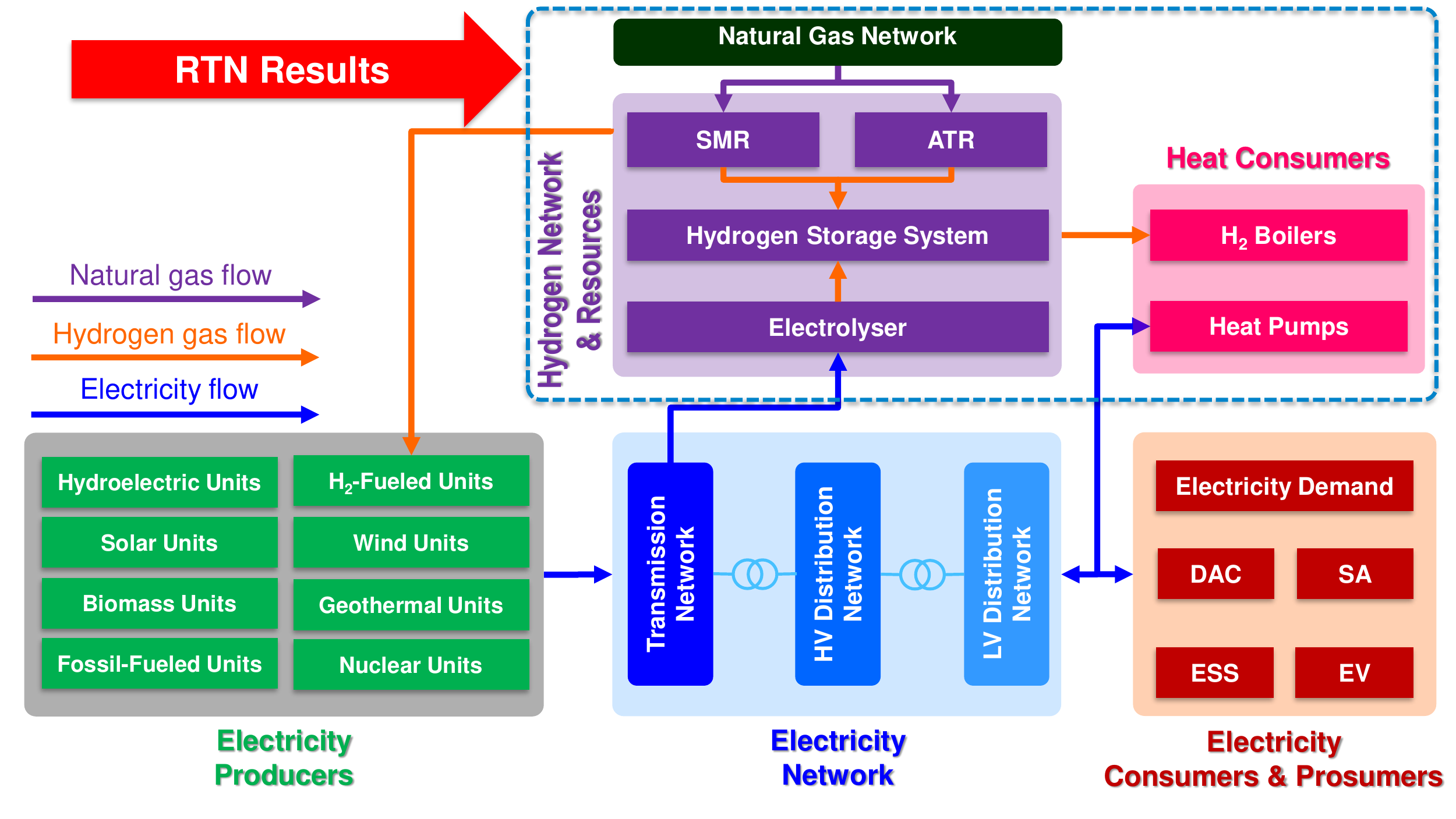}
	\caption{{\color{black}High-level diagram of WeSIM model}}
	\label{fig: WeSIM model}
\end{figure}

\subsection{Key features of WeSIM}
\label{subsec:WeSIMfeatures}
Capturing the interactions across different time scales and across different asset types is essential for the analysis of future low-carbon electricity systems that include flexible technologies such as energy storage and Demand Side Response (DSR). Deploying these technologies can improve not only the economics of real-time system operation, but also reduce the required investment into generation and network capacity in the long run.

In order to characterise these effects, and in particular trade-offs between different flexible technologies, it is critical that they are all included in a single integrated modelling framework. To this end, a comprehensive system analysis model WeSIM has been developed that is capable of simultaneously optimising long-term investment decisions against short-term operation decisions, across generation, transmission, storage and distribution infrastructures, in an integrated fashion. A detailed formulation of the model is provided in \cite{pudjianto2013whole}, and the model has been implemented in FICO Xpress Optimization framework \cite{FICO}.

WeSIM determines optimal decisions for investing into generation, network and storage capacity (both in terms of volume and location), aiming to supply the projected electricity demand in an economically optimal way, while at the same time ensuring appropriate security and adequacy levels for electricity supply. 

An advantage of WeSIM over most traditional models is that it is capable of simultaneously including system operation decisions and capacity additions to the system, with the ability to quantify trade-offs of using alternative mitigation measures, such as DSR and storage, for real-time balancing and transmission and distribution network and/or generation reinforcement management. 

Additionally, a key feature of the WeSIM model is the ability to optimally determine the necessary investments in distribution networks in order to meet demand growth and/or distributed generation uptake, based on the concept of statistically representative distribution networks \cite{gan2011statistical}.

It is essential to use high temporal and spatial granularity when studying electricity systems with high shares of variable renewables. Previous WeSIM applications have clearly demonstrated that in order to accurately quantify system operation and investment costs and assess $\mathrm{CO}_2$ emissions of various generation technologies, it is necessary to simulate second-by-second power balancing problems at the same time as multi-year investment decisions (e.g., low inertia in grids with high shares of renewables may trigger significant investment in flexible technologies). 

Also, electricity system decarbonisation will require capturing synergies and conflicts related to infrastructure requirements in local/district and national/trans-national levels, which is another essential capability of the WeSIM model.

WeSIM solves a single optimisation problem to find the set of least-cost investment and operation decisions, while considering two different time horizons: (1) short-term operation periods with a typical resolution of one hour or half an hour (while also taking into account frequency regulation and short-term reserve requirements), and (2) long-term investment, i.e., planning decisions with the time horizon of typically one year. 

All annual investment decisions and 8,760 hourly operation decisions are determined simultaneously in order to achieve the overall optimality of the solution. In summary, key features and constraints of WeSIM include: a) supply-demand balance, b) reserve and response requirements, c) generator operating limits, d) DSR capability, e) energy storage balance and operating limits, f) transmission and distribution network investment/reinforcement, g) carbon emission constraints, h) constraints on electricity imports/exports, and i) system adequacy and security constraints.

For the purpose of this paper and to enable the integration of WeSIM with RTN outputs, the original formulation of WeSIM has been extended to also explicitly consider technologies for hydrogen production, storage and transport. On the hydrogen production side these include electrolysis, ATR and SMR, all of which require electricity input, which affects the level of demand to be met by the power system. 

A high-level diagram of WeSIM (as used in this paper) is presented in Figure~\ref{fig: WeSIM model}.

\subsection{Linking the two models}
\label{subsec:ModelLinking}
\begin{figure}[t!] 
	\centering
    \includegraphics[width=\linewidth]{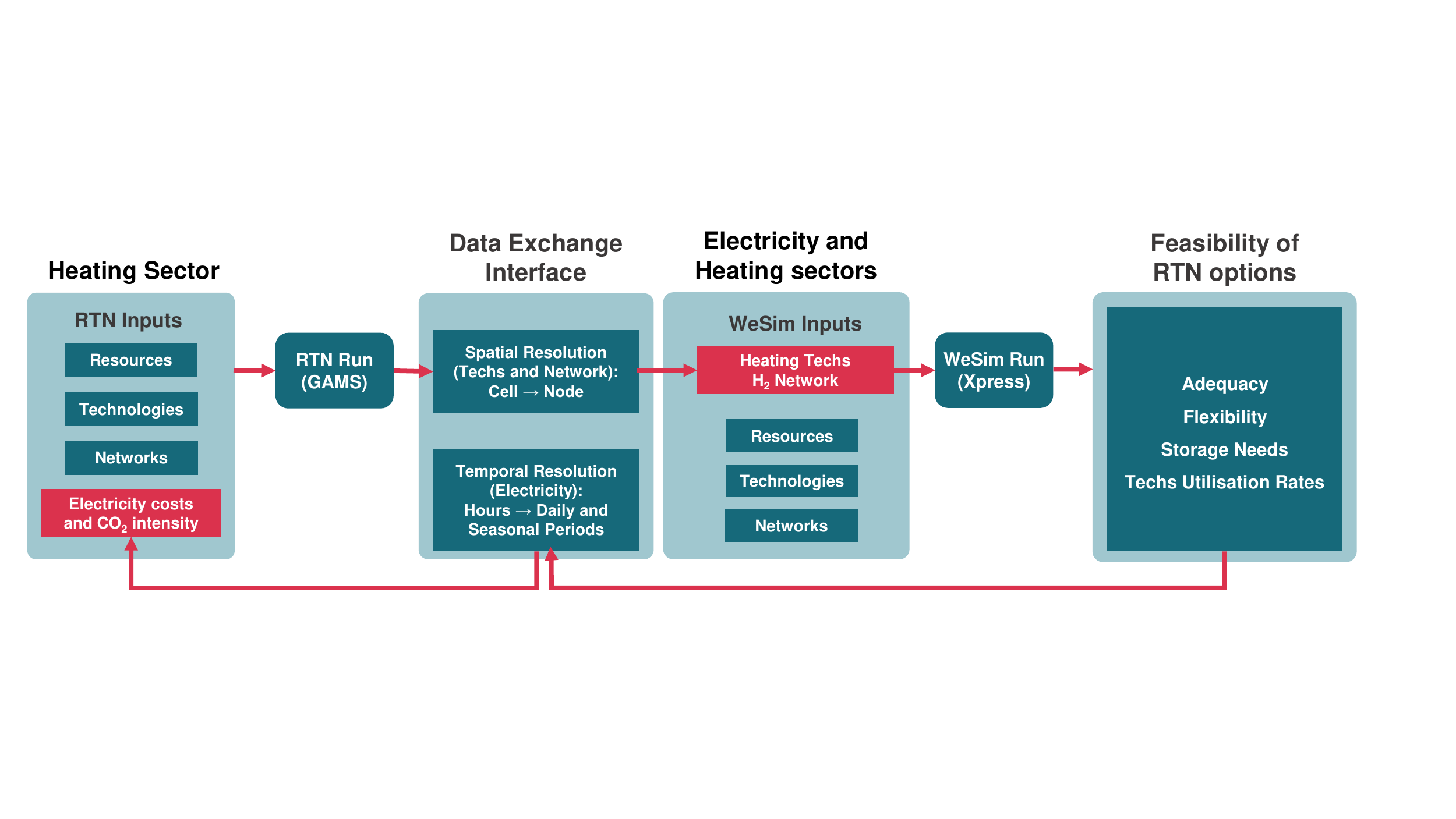}
	\caption{High-level diagram of interactions between RTN and WeSIM}
	\label{fig: RTN-WeSIM integration}
\end{figure}

Figure \ref{fig: RTN-WeSIM integration} presents the interactions between the two models used in this paper. RTN is used to find the cost-optimal investment plan to decarbonise the heat supply by solving a MILP problem with hourly electricity costs resulting from a first run of WeSIM, while WeSIM finds the optimal system configuration to decarbonise the electricity sector using a Linear Programming (LP) model formulation and using the heat technology mix and hydrogen network resulting from RTN.

In the RTN model, air-source heat pumps (ASHP), hydrogen boilers and hybrid ASHPs are considered as options to supply zero-carbon heat to consumers in the residential and small commercial heating sector. Hybrid ASHP corresponds to a combined unit of ASHP and hydrogen boiler, each contributing half of the total heat output capacity. 

Both ASHPs and hydrogen boilers are assumed to be zero-carbon heat technologies. However, the use of hydrogen boilers will be associated with some NO$_x$ emissions that could cause local air quality issues. These emissions are however expected to be lower than comparable NO$_x$ emissions from gas boilers. Under appropriate legislation the market should be able to deliver hydrogen boilers with high efficiency and minimal NO$_x$ emissions \cite{lewis2021optimising}, \cite{DNVGL2020}.

The focus of the paper is on the integration between the two models with the objective to identify cost-efficient zero-carbon heat pathways for individual end-user heating technologies. District Heat Networks (DHNs) were not included in the scope of the analysis, although it is understood that they could make a significant contribution to low-carbon heat sector \cite{millar2019district}.

One reason for not including DHNs is the increased complexity associated with aggregating and representing DHNs in energy system models due to the dependence of their cost and technical characteristics on network topologies and heat densities. Another reason is that, although it has been suggested in \cite{Zhang2018} that it would be cost-efficient to supply 16-21\% of UK’s heat demand from DHNs (predominantly in urban areas with sufficiently high heat density), the total system cost for a scenario with DHNs was found to be only marginally lower (by 0.4\%) than for a scenario that only considered hybrid HP solutions.

Similarly, a study by the UK Government \cite{HMGov2021DHN} suggested 20\% as an upper bound on the share of heat from DHNs, noting that the actual cost-efficient level is likely to be much lower. There is further uncertainty around the available volume of waste heat for supplying DHNs, as this volume may diminish in the future with the decarbonisation of energy-intensive industries \cite{manz2021decarbonizing}.

In earlier decades of the study period (i.e., in 2020s and 2030s) it is assumed that there will still be legacy conventional gas-fired boilers delivering heat, although they are gradually decommissioned due to the dual effect of reaching the end of lifetime and increasingly tighter carbon emission limit imposed on the heat supply. 

WeSIM on the other hand considers various types of zero-carbon (e.g., renewables, nuclear or hydrogen-fuelled generation), carbon-positive (e.g., gas-fired CCGT and CCS generation) and carbon-negative (e.g., direct air capture) technologies to supply electricity consumers with a specified level of carbon intensity. 

In addition to baseline electricity demand, the power sector configured by WeSIM also needs to supply the electrolyser and heat pump demand characterised in RTN (Figure \ref{fig: RTN model}), while the hydrogen production technologies (electrolysers, ATR and SMR) all supply hydrogen for both end-user hydrogen boilers as well as for any hydrogen-fuelled generators considered in WeSIM (Figure \ref{fig: WeSIM model}). Therefore, RTN and WeSIM models are co-dependent given that each RTN run requires inputs from WeSIM and vice versa.

The iterative procedure illustrated in Figure \ref{fig: RTN-WeSIM integration} starts with the first set of WeSIM runs for the three snapshot years (representing the three decades, 2030-2040, 2040-2050 and 2050-2060) while assuming a fully electrified GB heat supply i.e., with all domestic and commercial heat requirements delivered through heat pumps and no contribution from hydrogen to zero-carbon heating. 

These runs determine the least-cost configurations of the electricity system for each of the three representative years. They also produce hourly series of electricity prices (estimated using dual values, i.e., marginal electricity prices determined by the model) for the three representative years. 

Hourly prices are then aggregated into 16 electricity price levels according to the RTN time slice specifications. RTN uses this input to determine the cost-optimal decarbonisation trajectory for the heat sector i.e. the optimal mix of hydrogen and electricity in the heat supply mix given the variations in electricity prices provided by WeSIM. 

This optimal investment plan is passed onto WeSIM, including capacities and locations of hydrogen production technologies (electrolysers, ATRs and SMRs), hydrogen transport infrastructure and hydrogen storage systems, as well as the mix of end-use supply of domestic and commercial heat through heat pumps and hydrogen boilers. 

Based on this information, the electricity and hydrogen demand profiles in WeSIM are updated, and the next iteration of the model is run to find the updated cost-optimal system configuration for the electricity sector. This solution also includes an updated set of hourly electricity prices for RTN. In principle, this iterative procedure could continue until there is little or no change between solutions for two successive iterations. For the sake of simplicity, only the results of the first two iterations are presented in this paper.

In its original form, the RTN solution specifies the infrastructure and resources required to satisfy the residential and commercial heating demand in each of the 51 cells. Before passing that information onto WeSIM, the cell-level solutions need to be aggregated into 5 regions used in WeSIM to represent the GB electricity system based on a predefined cell-to-region mapping procedure. The aggregated information is then sent to WeSIM to re-optimise the capacity mix and operation of the electricity system as well as the hourly provision of heating based on capacities determined by RTN.

\subsection{System scenarios and main assumptions}
\label{subsec:Scenarios}
System-level assumptions used in the paper, and in particular the assumptions on the heat demand level, were broadly based on the System Transformation scenario from National Grid’s Future Energy Scenarios (FES) \cite{NGFES2020}. 

To study the trajectory of GB heat decarbonisation, three decades have been considered in the analysis: 2030-2040, 2040-2050 and 2050-2060. In the last decade (2050-2060) it was assumed that heat supply needs to be fully decarbonised i.e., producing zero $\mathrm{CO}_2$ emissions; therefore the only options to deliver zero-carbon end-use heat in RTN were electric heat pumps, hydrogen boilers or their hybrid combination. 

Carbon emission limit from heat supply was reduced linearly from its 2020 level to zero in 2050-2060, taking into account emissions from using natural gas in reforming technologies and carbon intensity of electricity used to run electrolysers. When determining the cost-efficient mix of producing hydrogen for heat supply under a carbon constraint, the RTN model had an option to invest in negative-emission technologies such as biomass gasification with CCS in order to offset any residual emissions from SMR and ATR facilities.

Carbon intensity of the power system in WeSIM was constrained to 41~$\mathrm{gCO_2}$ per kWh in 2030-2040, and then to net zero carbon emissions in 2040-2050 and 2050-2060.

The wholesale price of natural gas used in reformers in RTN was assumed to follow a seasonal variation, with the price of £15.81/MWh in summer and £17.71/MWh in winter. Retail electricity prices used in RTN were obtained from WeSIM wholesale prices by imposing a ratio of 2.2 between wholesale and retail price and a retail price cap of £528/MWh.

Costs of electricity supply technologies used in WeSIM were assumed in line with \cite{IDLES2021}. Cost assumptions for hydrogen production, storage and transportation technologies in RTN were taken from \cite{sunny2020needed}, while their future cost reduction trends were based on the recent UK Government study on the cost of hydrogen \cite{BEIS2021H2}. 

A build rate limit of 8~GW of new hydrogen production capacity per year was imposed in RTN to match the build rates implied in Climate Change Committee's Sixth Carbon Budget \cite{CCC2020SixthCB}. Total annual heat demand for combined residential and commercial sectors assumed in the model was assumed to reduce linearly from 541~TWh in 2020 to 476~TWh in 2050-2060, reflecting improvements in energy efficiency, i.e., building insulation levels. 

Both RTN and WeSIM assumed seasonal variations in the Coefficient of Performance (COP) of ASHPs in line with typical variations in outdoor temperature in GB. In line with the wider decarbonisation objectives, the electricity system was also assumed to incorporate a high share of electrified road transport (i.e., electric vehicles).

Interconnection between GB and continental Europe was also modelled in WeSIM, assuming energy neutrality for the GB electricity system (i.e., although at any hour interconnections can be used to export or import electricity, the total exports match total imports over the course of a year). Security of the GB power system was ensured by enforcing a standard Loss of Load Expectation (LOLE) criterion of up to 3~hours per year.

Further details on the technical and cost assumptions for various technologies used in this study are provided in the Appendix.

\section{Results and discussion}
\label{sec:ResultsDiscussion}
This section outlines the key results obtained from running the two models, focusing on how their interaction affects the proposed power system configuration and the heat supply mix. The results of this analysis are highly dependent on the input assumptions considered, which are subject to a large degree of uncertainty. Therefore, the main focus is on how the two models could effectively exchange information and in the process gradually improve the quality of the solution. 

\subsection{First WeSIM run}
\label{subsec:FirstWeSIMrun}
In the first instance, WeSIM is run to cost-optimise the electricity generation mix under system-wide carbon constraints specified in Section~\ref{subsec:Scenarios}. As mentioned before, this optimisation was carried out under the assumption that the domestic and commercial heating sector is 100\% electrified, which is obviously not a realistic option, especially in the earlier decades in the analysed period. Nevertheless, this approach was deemed justified for the first WeSIM iteration in order to test the boundaries of the solution space before the solution is refined through subsequent iterations, also considering that part of the renewable electricity consumed for heating in the first WeSIM run could be instead used for hydrogen generation to supply heat in the RTN optimisation.

\begin{figure}[t!] 
	\centering
    \includegraphics[width=\linewidth]{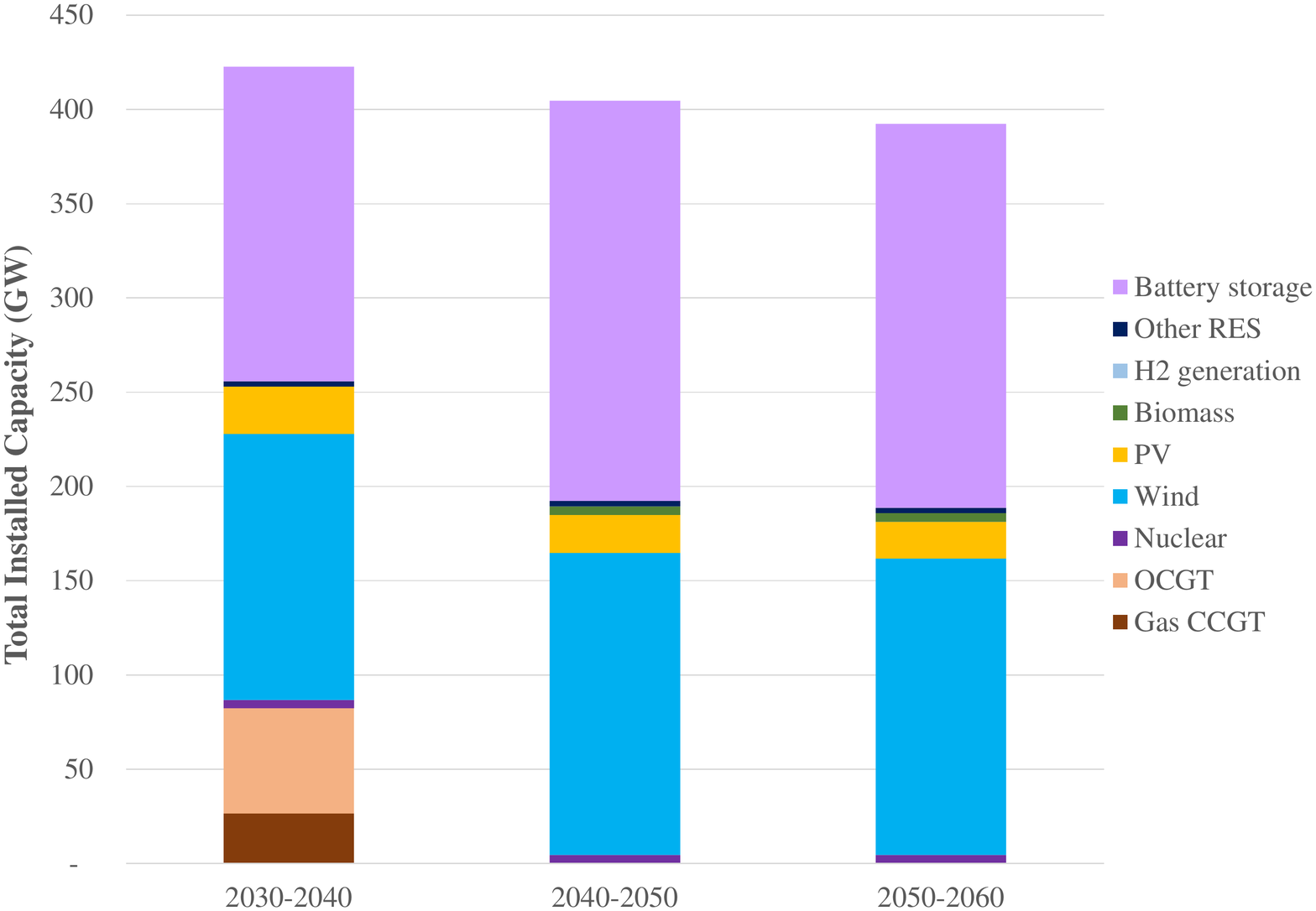}
	\caption{Total electricity generation and battery storage capacity in first WeSIM run}
	\label{fig: WeSIM_Run1}
\end{figure}

As shown in Figure~\ref{fig: WeSIM_Run1}, the model builds a portfolio of low-carbon generation technologies to meet a very high level of demand due to the assumption on full electrification of domestic and commercial heating. The total installed generation capacity gradually reduces between the first and last decades as the level of heat demand decreases due to improved energy efficiency. 

The low-carbon generation technologies chosen for installation in the GB system mainly include onshore and offshore wind and solar PV, supported by a very large volume of battery storage capacity (between about 150 and 200~GW). Other technologies in the generation mix include nuclear, biomass and other RES, while in the first decade there is also a significant amount of gas generation, which is able to make a contribution to electricity supply because of a less stringent carbon constraint. 

It is clear that delivering such a high volume of renewable and battery storage capacity by the first decade would not be feasible, and neither would it be realistic to fully electrify the domestic and commercial heat supply in such a short time frame. Still, it is useful to explore the boundaries of the system and establish what would be the composition of an unconstrained system portfolio.

\begin{figure}[t!] 
	\centering
    \includegraphics[width=\linewidth]{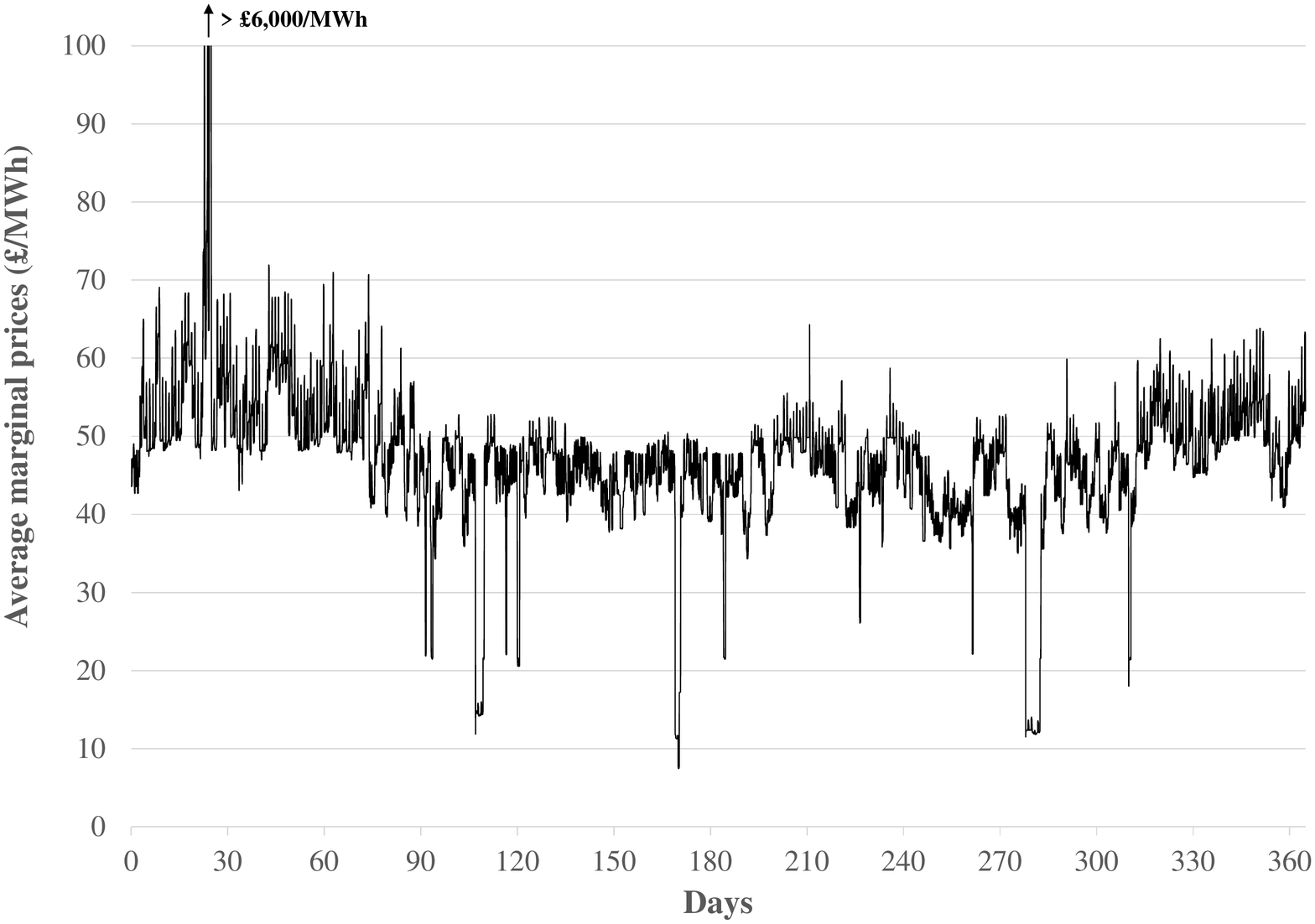}
	\caption{Hourly electricity prices from first WeSIM run in 2050}
	\label{fig: WeSIM_prices}
\end{figure}

WeSIM also produces a set of hourly electricity prices that reflect short-term fluctuations in supply and demand balance in the system. Figure~\ref{fig: WeSIM_prices} shows the hourly power price variations obtained from WeSIM for year 2050. Electricity price broadly varies in the range £50-70/MWh during winter, and between £40-50/MWh during summer, with occasional price drops to lower levels during periods of high renewable output. 

Around Day 25 of the year, the cold temperatures cause a spike in electricity demand for heating (also accentuated by lower COP values), and as a consequence also drives a spike in electricity prices to over £6,000/MWh (note that these are outside of scale in  Figure~\ref{fig: WeSIM_prices} as they would dwarf other price variations). 

Marginal prices determined in WeSIM reflect the scarcity of electricity supply not just in terms of energy but also in terms of necessary marginal investment in generation, storage and network infrastructure required to meet high peak demand driven by extremely cold weather. Therefore the prices associated with very high peak demand around day 25 tend to be well above £1,000/MWh.

Hourly prices from WeSIM are aggregated into the 16 time slices used to represent each year in RTN, by finding the average price for all hours that are associated with a given time slice. As discussed earlier, 4 representative seasons are used: winter (consisting of 124 days), autumn/spring (85 days), summer (155 days), and an explicitly represented winter peak day. Within each day four slices were used to represent night (0-7h), day (7-17h), peak (17-20h) and evening (20h-midnight). 

In addition to temporal aggregation, the wholesale prices from WeSIM were further converted to retail prices before being used in RTN, by applying: a)~retail to wholesale price ratio of 2.2, and b)~retail price cap of £528/MWh.

\begin{figure}[t!] 
	\centering
    \includegraphics[width=\linewidth]{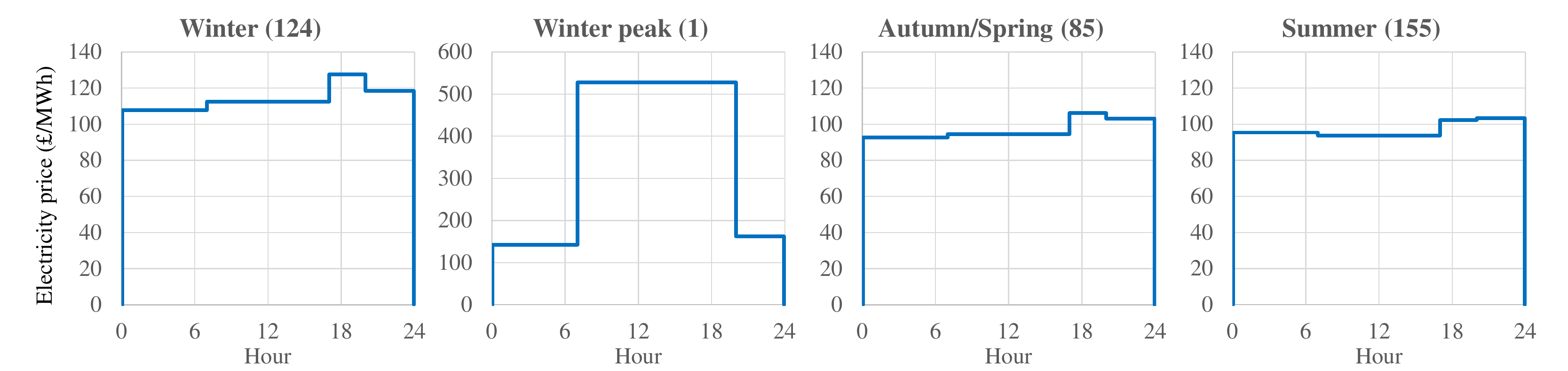}
	\caption{Retail electricity prices across time slices in RTN in 2050. Numbers in brackets denote the number of days in each representative season.}
	\label{fig: RTN_prices}
\end{figure}

Figure~\ref{fig: RTN_prices} illustrates the retail electricity prices used in RTN across various time slices in 2050, obtained based on WeSIM wholesale hourly prices in Figure~\ref{fig: WeSIM_prices}. Note that the scale used for the winter peak day is much larger than for the other three representative days, given that for two time slices on winter peak day the price hits the retail price cap level.

\subsection{First RTN run} \label{subsec:FirstRTNrun}
\begin{figure}[t!] 
	\centering
    \includegraphics[width=\linewidth]{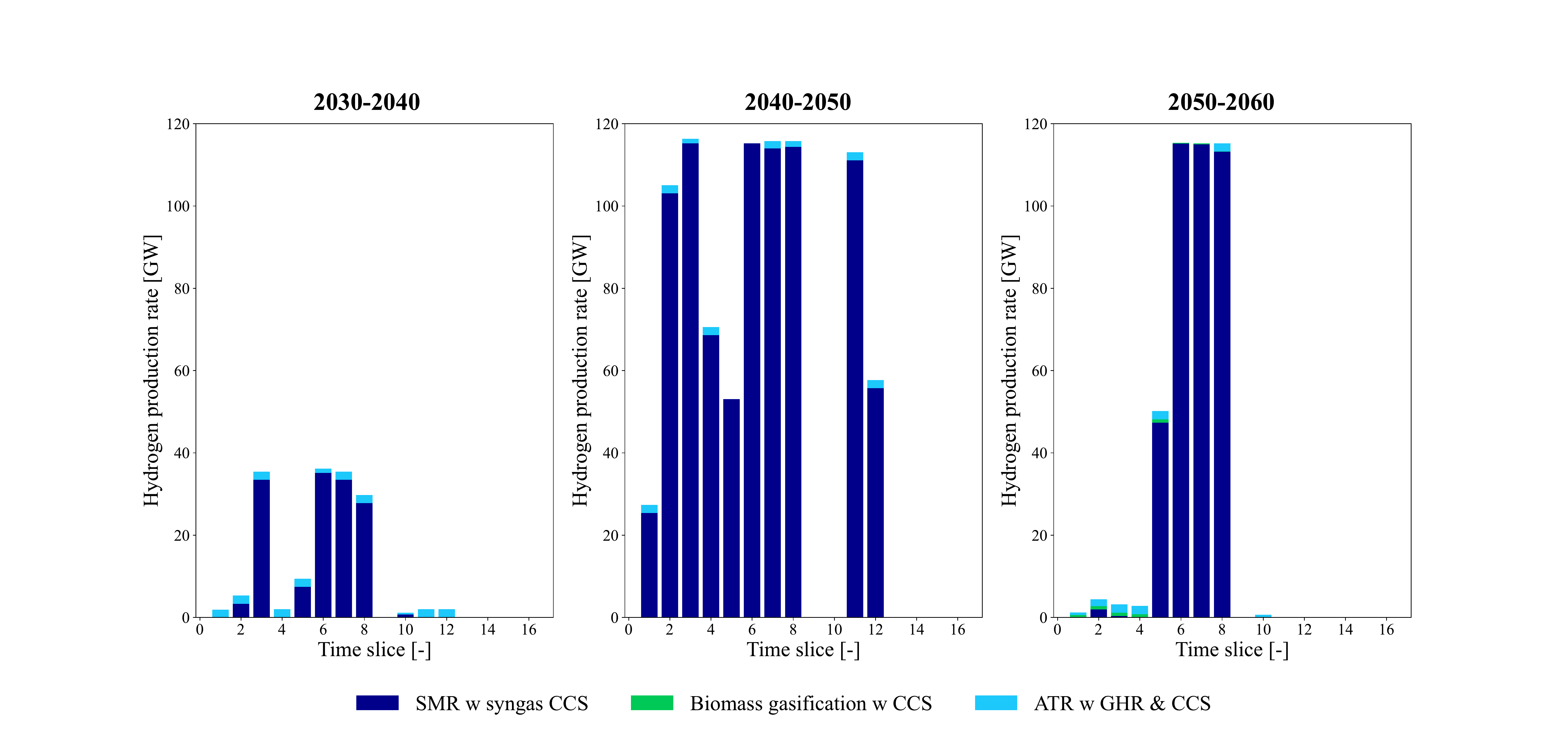}
	\caption{Hydrogen supply in first RTN run across three decades}
	\label{fig: RTN_H2_1st}
\end{figure}

The first iteration of RTN uses the retail electricity prices obtained as described in the previous section to determine the cost-optimal split of low-carbon heat supply between electricity and hydrogen and also to determine the required capacities of hydrogen production, storage and transport infrastructure. Figure~\ref{fig: RTN_H2_1st} presents the resulting mix of hydrogen supply technologies across the three decades and for different time slices.

Total hydrogen generation capacity in 2030-2040 is about 35~GW, increasing to about 115~GW in 2040-2050. The model therefore deploys as much hydrogen production capacity as allowed by the assumed national build rate limit of 8~GW/year. The preferred hydrogen generation technology is SMR with syngas capture due to its relatively low investment cost although it is less efficient and has higher emissions than the alternatives considered in the model. 

ATR with Gas Heated Reformer (GHR) and CCS is only deployed in cells with available underground hydrogen storage. In 2050-2060, 0.8~GW of biomass gasification capacity is added as a net negative emission technology, in order to compensate for the emissions from SMR plant and meet the net zero carbon constraint. 

To minimise the need for transport infrastructure, most of the hydrogen requirements are met locally except in the Greater London area, where heat demand exceeds the national limit on build rate and makes it necessary to build transport infrastructure between Central and South East England. The majority of the CCS pipeline system in England is built by 2030-2040 to transport $\mathrm{CO}_2$ from hydrogen production to injection wells in the East Irish Sea and North Sea.

The net zero carbon constraint imposed in 2050-2060 requires offsetting of any $\mathrm{CO}_2$ emissions from hydrogen production or producing green hydrogen via electrolysis. However, electrolysis does not seem to be cost-competitive to other sources of hydrogen. Negative emission technologies require higher investment, which makes the use of large volumes of hydrogen throughout the year expensive. 

Therefore, producing hydrogen from SMR while at the same time offsetting emissions via biomass gasification is only cost-effective for the winter peak demand day. In order to compensate for SMR emissions, biomass gasification plants need to operate not only during the winter peak demand day but also during other winter days to ensure the appropriate volume of carbon emissions is captured.

The results suggest that the volume of hydrogen used in the heating sector would be the greatest in the intermediate time horizon (2040-2050), as a means to transition the heat sector from natural gas-based to predominantly electricity-based supply in a bid to eliminate carbon emissions. Even in the fully decarbonised heat sector in 2050-2060, hydrogen still plays a role as source of heat during peak demand periods, which is preferred to installing additional capital-intensive ASHP capacity that would operate at low utilisation factors.

\begin{figure}[t!] 
	\centering
    \includegraphics[width=\linewidth]{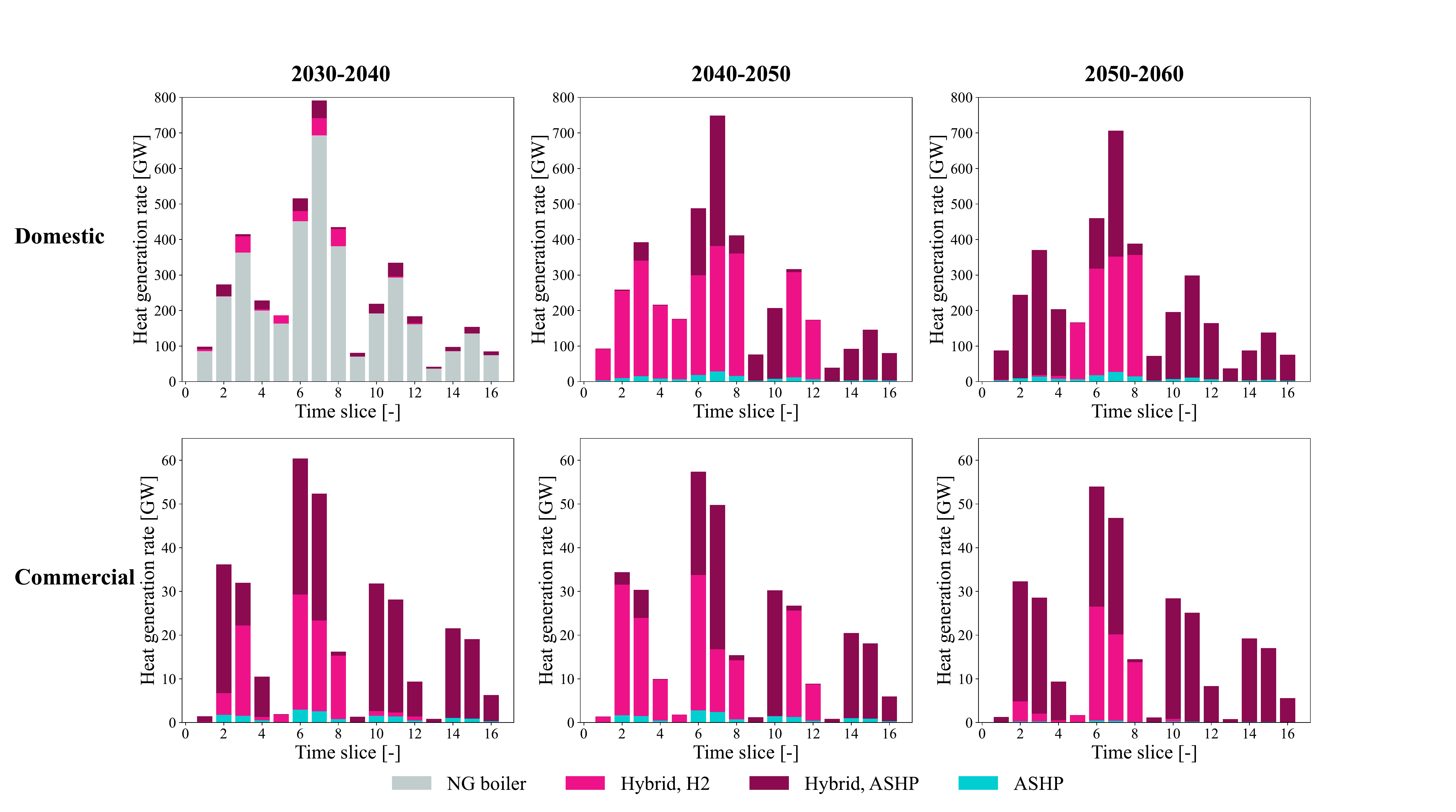}
	\caption{Supply of domestic and commercial heat in first RTN run across three decades}
	\label{fig: RTN_Heat_1st}
\end{figure}

The cost-optimal mix of heat supply technologies found in the first iteration of RTN is shown in Figure~\ref{fig: RTN_Heat_1st}. The extent to which hydrogen and electricity are used is driven by the emission reduction trajectory, allowing natural gas boilers that are in place today to operate until 2040. 

Commercial demand is decarbonised earlier as its volume is assumed to be considerably lower and its heat demand profile less peaky than the domestic demand. By 2050-2060, the $\mathrm{CO}_2$-neutral production of hydrogen via renewable electricity and electrolysis is not cost-competitive to the direct use of electricity in ASHP. ASHP are favoured due to the higher overall efficiency and the retail price cap assumed to be in place to protect domestic and commercial customers from spikes in wholesale electricity prices.

Hydrogen is first used in 2030-2040 to contribute to meeting the heat demand of the commercial sector during winter peaks. A small share of domestic demand also transitions to hybrid hydrogen-electric systems. 

In 2040-2050 the share of hydrogen in commercial heating increases further, supplying the majority of heat demand during winter and autumn/spring seasons; a similar supply pattern is observed for the domestic sector. In 2050-2060, no new hydrogen production capacity is added as hydrogen is only used to top up commercial and domestic heat during the winter peak day, while at other times hybrid ASHP units operate in electric mode to cover the entire heat demand. 

Conventional ASHP units are deployed to a small extent in areas where hydrogen supply is not cost-effective due to remoteness and relatively low heat demand. The reason for the reduction in the share of hydrogen in the low-carbon heat supply in 2050-2060 compared to the previous decade lies in the net-zero carbon emission limit in 2050-2060, which makes hydrogen production from reformers with non-zero $\mathrm{CO}_2$ emissions relatively less attractive given that their carbon emissions need to be offset by installing net carbon negative technologies such as biomass gasification.

\subsection{Second WeSIM run}
\label{subsec:SecondWeSIMrun}
\begin{figure}[t!] 
	\centering
    \includegraphics[width=\linewidth]{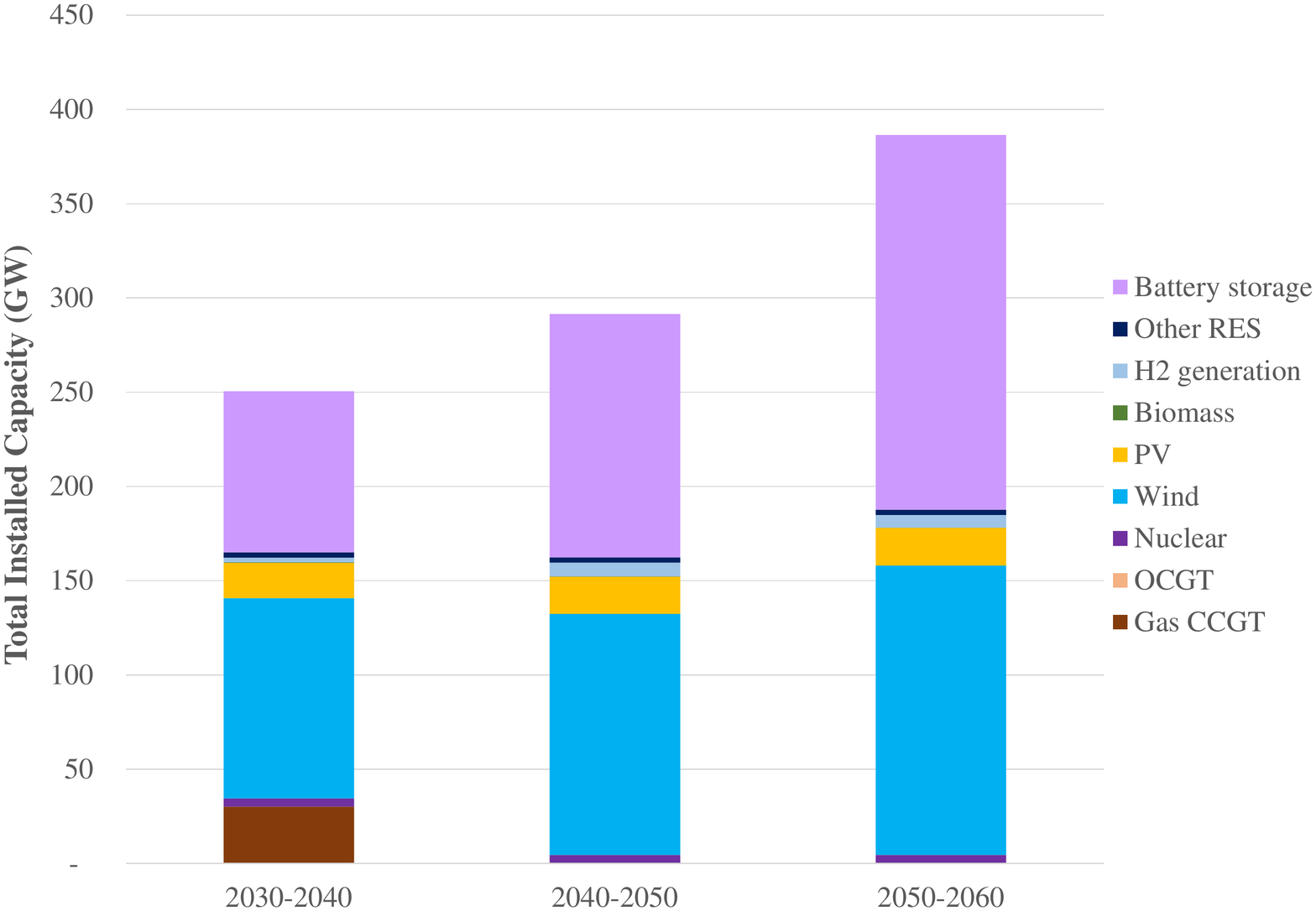}
	\caption{Total electricity generation and battery storage capacity in second WeSIM run}
	\label{fig: WeSIMRun2}
\end{figure}

The second run of WeSIM used the domestic and commercial heat demand split between electricity and hydrogen as determined in first RTN run. The resulting capacity mix in the second WeSIM run is presented in Figure~\ref{fig: WeSIMRun2}. 

When compared with the results of the first WeSIM run (Figure~\ref{fig: WeSIM_Run1}), it is immediately obvious there are less extreme requirements for generation and storage capacity in the short-run (although the required capacity additions to today's system are still very ambitious). Instead, one can observe a more gradual increase in supply over the decades as the level of heat electrification rises in line with the assumed carbon reduction trajectory for heating imposed in RTN. 

As opposed to the first run, no peaking OCGT capacity is required in 2030-2040 as the actual share of electrified heating is relatively low. The total combined capacity of wind and solar PV in 2030-2040 reduces from 165~GW in the first run to 125~GW, while battery storage capacity effectively halves from 167 to 85~GW.

The capacity mix in 2050-2060 is very similar to the one found in the first run (Figure~\ref{fig: WeSIM_Run1}), which assumed 100\% heat electrification. The main difference is that the requirements for firm low-carbon generation, which in the first run were met by nuclear and biomass generation, are now met by nuclear and hydrogen generation. This is enabled by the opportunity to cost-effectively produce hydrogen (mostly through SMR capacity featuring in the RTN solution), some of which is used to produce electricity in hydrogen-fuelled CCGT and OCGT units during times when firm low-carbon output is required in the system.

Driven by the lower electricity demand, the updated electricity prices for the first decade (2030-2040) were also considerably lower; as an illustration, the (non-weighted) average price reduced from £52.2/MWh to £48.3/MWh. The differences for the other two decades were much smaller given the smaller differences in electrified heat demand between the first and second run. The first run average prices of £54.3/MWh observed in both 2040-2050 and 2050-2060 reduced to £53.8/MWh and £54.2/MWh, respectively, in the second run.

The total volume of wind generation proposed by the model reaches around 150~GW in 2050-2060, which is significantly higher than the currently installed wind capacity in the UK (24.6~GW as of April 2022). Nevertheless, this capacity is still in line with the estimated wind potential in the UK. For instance, National Grid’s Future Energy Scenarios project the wind generation capacity in 2050 at 158~GW \cite{NGFES2020}, while the Climate Change Committee’s Sixth Carbon Budget reports an estimate of 124-341~GW for the UK’s wind potential \cite{CCC2020SixthCB}. Annual generation output across different technologies for both WeSIM runs is provided in Table~\ref{table: Annual_el_production} in the Appendix.

\subsection{Second RTN run} \label{subsec:2ndRTNrun}
\begin{figure}[t!] 
	\centering
    \includegraphics[width=\linewidth]{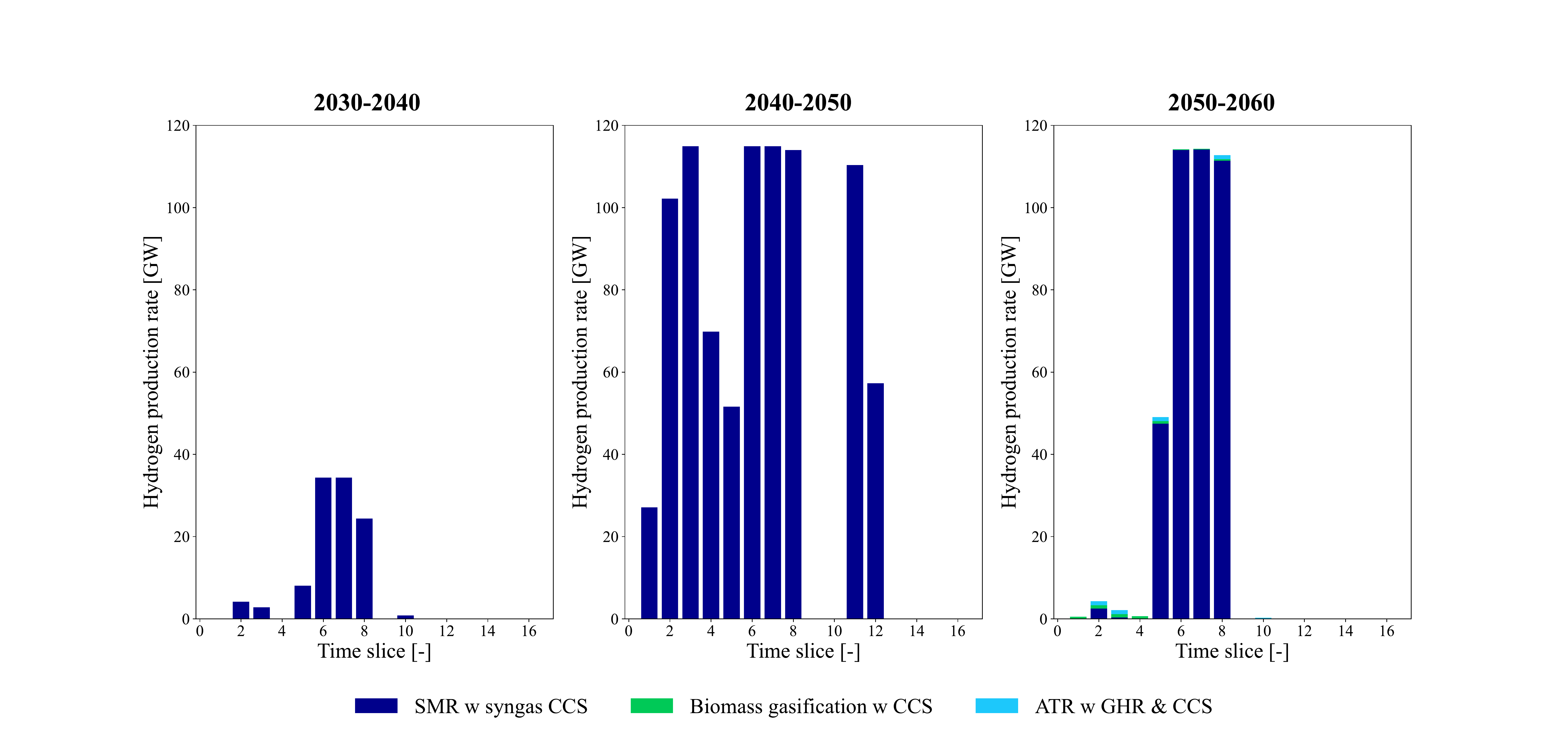}
	\caption{Hydrogen supply in second RTN run across three decades}
	\label{fig: RTN_H2_2nd}
\end{figure}

An updated set of electricity prices produced in the second WeSIM run (similar to the one in Figure~\ref{fig: WeSIM_prices}) was used for the second RTN run, after applying the same retail to wholesale price ratio and a retail price cap to obtain time slice prices. The resulting hydrogen production mix across different decades and time slices is shown in Figure~\ref{fig: RTN_H2_2nd}.

There are no significant changes in total hydrogen production capacity between the first RTN run (Figure~\ref{fig: RTN_H2_1st}) and the second. About 80~GW of capacity is still added between 2030-2040 and 2040-2050, corresponding to the maximum allowed build rate for the 10-year interval. The main change from the first iteration is that no ATR capacity is added until 2050-2060, as it is made less attractive by lower electricity prices, particularly during 2030-2040, when most ATR investment decisions were made in the first RTN run.

The pattern of hydrogen production remains very similar for the last two decades. On the other hand, the output pattern in 2030-2040 changes visibly as much less hydrogen is used to supply heat during the winter day (time slices 1-4) as the electricity is available at lower prices than in the first iteration and therefore ASHPs take over most of the heat supply for that day. 

Also, there is no more hydrogen output from ATR that was previously used locally in some cells during the winter. Similarly, in the spring/autumn season the contribution of hydrogen to heat supply becomes negligible. Hydrogen remains to be used at a high rate during the winter peak day in 2030-2040, when electricity prices hit the retail cap level (time slices 5-8).

\begin{figure}[t!] 
	\centering
    \includegraphics[width=\linewidth]{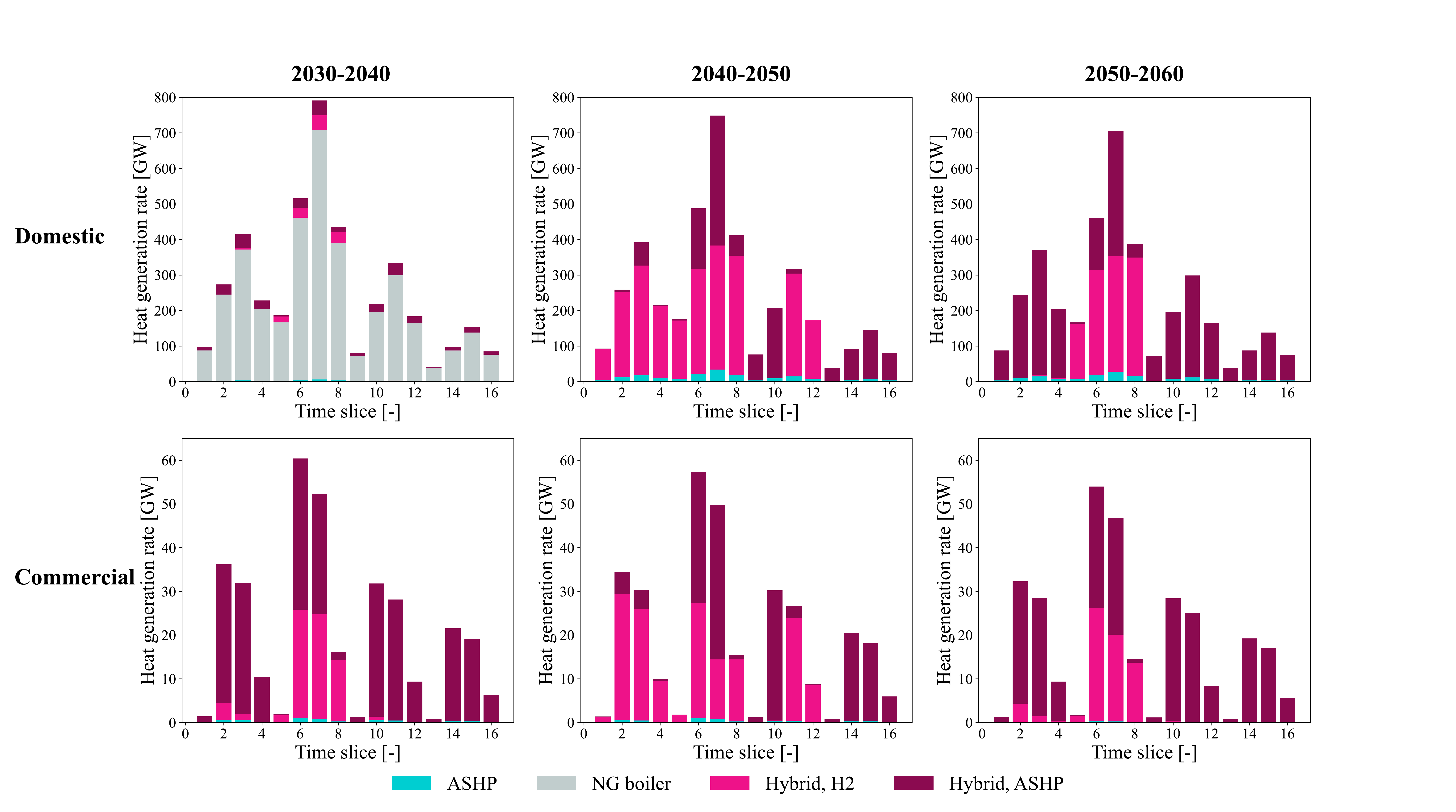}
	\caption{Supply of domestic and commercial heat in second RTN run across three decades}
	\label{fig: RTN_Heat_2nd}
\end{figure}

As shown in Figure~\ref{fig: RTN_Heat_2nd}, the structure of end-use heat supply does not change significantly from the first RTN run (Figure~\ref{fig: RTN_Heat_1st}), with high presence of natural gas in the domestic sector and a mix between electricity and hydrogen in the commercial sector in the first decade. Hydrogen is still the dominant source of heat during 2040-2050, while in the final decade the net-zero carbon target makes ASHPs the preferred heat source and hydrogen takes over the role of top-up heat source during extremely cold periods. 

The main notable difference between the two RTN runs is the lower use of hydrogen in winter days in 2030-2040 (time slices 1-4), which is driven by lower electricity prices used for this decade in the second RTN run, as discussed earlier. In both sectors, hybrid ASHPs are the dominant option as they have the advantage of supplementing the heat supplied from electricity with heat generated using the hydrogen boiler sub-unit during extreme days.

The volume of natural gas used for heating, whether directly in gas boilers or indirectly by producing hydrogen in reformers, declines over the study period. The estimate used for the current annual gas consumption is 580~TWh. In 2030-2040, the consumption of gas for heating drops to 368~TWh, as the combined result of energy efficiency improvements and partial electrification of heating; in the same decade, about 10~TWh of gas is used for hydrogen production in reforming plants. In the following decade (2040-2050) the use of gas for direct heating reduces to zero due to a more stringent emission target, whereas 379~TWh of gas is now used for producing hydrogen. Finally, in 2050-2060, as the heating sector gets predominantly electrified, the use of gas for hydrogen production drops to just 10~TWh.

\subsection{Geographical distribution of hydrogen and CCS infrastructure}
\label{subsec:GeoDistribution}
\begin{figure}[t!] 
	\centering
    \includegraphics[width=\linewidth]{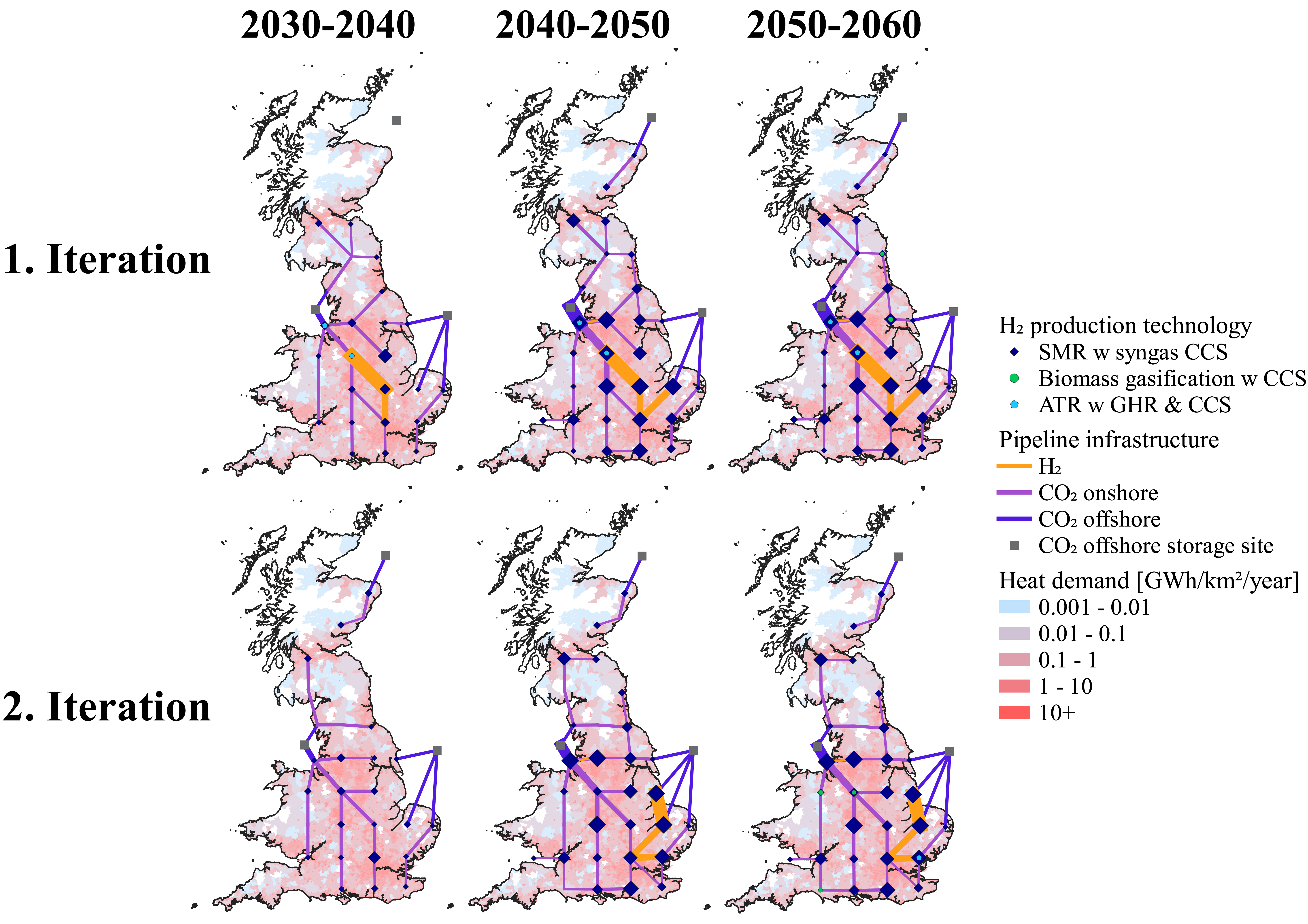}
	\caption{Geographical distribution of infrastructure for hydrogen production and transport, and for $\mathrm{CO}_2$ transport and storage in RTN across three decades and for both iterations. Sizes of hydrogen sources and widths of pipelines are proportional to their capacity. Plots are shown on top of GB heat density map.}
	\label{fig: RTN_GBmap}
\end{figure}

In addition to sizing the hydrogen production capacities, RTN also sizes the hydrogen and $\mathrm{CO}_2$ transport network across its 51 cells used to represent the GB system. Figure~\ref{fig: RTN_GBmap} provides more detail on the geographical configuration of these two pipeline systems, showing the locations and sizes of hydrogen production plants across different cells along with the locations and sizes of hydrogen and $\mathrm{CO}_2$ pipeline infrastructure for all three decades and both RTN runs. Note that the sizes of hydrogen sources and widths of pipelines are plotted proportional to their capacity. All plots are superimposed over the GB heat density map.

In the first RTN iteration hydrogen pipeline capacity is built already in the first decade (2030-2040) to connect Central and South East England and supply Greater London area, while also benefiting from cavern hydrogen storage in the Cheshire area. The CCS pipeline system also gets built by 2030-2040, covering most of England, in order to transport the carbon captured in the process of reforming gas to produce hydrogen. 

The injection wells in the East Irish Sea and Southern North Sea are accessed first, followed by the Northern North Sea storage site. The extensive onshore $\mathrm{CO}_2$ pipeline network is mostly built to transport $\mathrm{CO}_2$ from South England to the East Irish Sea, which is preferred due to the proximity to the shore and therefore lower offshore pipeline cost. The storage in the Southern North Sea is mostly used for areas close to its shore.

As discussed before, the role of hydrogen in 2030-2040 diminishes in the second RTN iteration due to lower electricity prices, affecting the build schedule and configuration of hydrogen network to supply the London area. In the first iteration with higher electricity prices, the network design is driven by the co-location of ATR and intra-day hydrogen storage, which combine to produce hydrogen from ATR using cheaper electricity during the night (note that ATR uses more electricity than SMR per unit of hydrogen output), and using storage to supply hydrogen to the London area. 

In the second RTN run with lower electricity prices in 2030-2040, the ATR is no longer chosen as hydrogen source, so the location of hydrogen sources and pipelines is driven by the proximity to offshore $\mathrm{CO}_2$ storage on the eastern GB shore. The layout of the $\mathrm{CO}_2$ onshore pipeline network changes as well, in particular around the English-Scottish border and the Dorset area in the southwest.

This analysis did not consider the possible growth in hydrogen demand fr industry and transport as these sectors decarbonise. Including these sectors in the analysis would result in higher capacity requirements for hydrogen production and transport (including the CO$_2$ transport). Given that the UK’s energy-intensive industries and major shipping ports are concentrated around the coast of north England, south Wales and Scotland, new hydrogen production and stronger transport capacity would be required in those regions. 

This analysis did not consider any additional hydrogen demands that could arise from the decarbonisation of the industry and transport sector in the future. Extending the analysis by including these sectors would result in higher capacity requirements for hydrogen production and transport (including the CO$_2$ transport). Demand from major shipping ports and the energy-intensive industries at the coast of north England, south Wales and Scotland would introduce the most changes to the CO$_2$ pipeline infrastructure as the capacities of pipelines connecting these locations to offshore sites would need to be added or expanded. Assuming local hydrogen production would still be preferred, no significant changes to the hydrogen pipeline network would be expected.

Additional hydrogen demand for aviation and heavy goods transport would not significantly change the system configuration proposed by Figure~\ref{fig: RTN_GBmap} given that the main hubs for aviation and goods transport are well aligned with the main heat demand centres.

\section{Conclusion}
\label{sec:Conclusion}
This paper has proposed an approach for soft-linking two energy system models, RTN and WeSIM, in order to utilise their respective strengths to study the multifaceted issue of decarbonising GB heat supply by 2050. This dual-model approach allowed for a technology-rich representation of hydrogen production, storage and transport options with high spatial granularity provided by RTN, while at the same time taking advantage of high-fidelity representation of the low-carbon power system in the WeSIM model, using a fine temporal resolution and high level of technical detail. 

The proposed approach allowed for the direct representation of linkages between the power system and the provision of end-use heat using hydrogen and electricity. Key input from WeSIM into RTN were hourly electricity prices that reflected the long-run marginal cost of providing electricity from carbon-constrained electricity system. 

RTN used these inputs to cost-optimise the portfolio of zero-carbon end-use heat technologies as well as the portfolio of hydrogen production, storage and transport technologies along with the CCS infrastructure. The split of end-use heat supply between electricity and hydrogen was then returned to WeSIM to update its electricity demand profiles in the next iteration and provide an updated set of electricity prices to RTN for the next iteration.

Iterating between the two models led to gradual improvements of the solution, even when the process started with an extreme assumption of 100\% electrification in the initial WeSIM run. This assumption resulted in overestimated requirements for new electricity generation capacity as well as higher electricity prices; however, with the revised electrified heating demand provided by RTN as input to the second iteration, WeSIM produced more moderate requirements for generation capacity as well as reduced electricity prices.

Quantitative results obtained by iterating between the two models suggest that their soft-linking can improve the quality of the low-carbon heating solution. Key findings based on the iterative use of the two models can be summarised as follows:

\begin{itemize}
    \item Hydrogen can play a significant role in transition towards full heat decarbonisation, especially in the medium term, due to its ability to provide low-carbon heat with relatively low investment (both at end-use and large-scale production level) and operation costs. However, due to the non-zero carbon emissions, the attractiveness of gas reforming technologies even when equipped with CCS reduces as the system evolves towards the net-zero carbon constraint. At the same time, producing hydrogen from electrolysis does not appear cost-competitive compared to direct use of electricity in ASHPs. Nevertheless, even with a net-zero carbon heating supply hydrogen from SMR with CCS remains a cost-efficient option for providing peak heat supply capacity, in order to cope with rare but extreme cold weather events.
    \item In the short term (2030-2040) the majority of heat in the domestic sector continues to be provided by natural gas. The commercial sector, with lower demand levels and a less peaky profile, is decarbonised using a mix of electricity and hydrogen a decade earlier than the domestic sector.
    \item Model outputs suggest that the most cost-effective hydrogen production technology is SMR with CCS, despite a lower efficiency and higher carbon intensity than ATR or electrolysers. Only a small amount of ATR capacity is installed, while in the net-zero system in 2050-2060 the carbon emission from SMR need to be offset by installing a small amount of biomass gasification plant with net negative carbon emissions.
    \item In order to meet the ambitious carbon reduction targets in the electricity sector (which drop to net zero already in 2040-2050), it will be necessary to significantly expand renewable generation and in particular the offshore wind in the UK. To manage fluctuations in renewable output it will be necessary to ensure adequate levels of system flexibility by installing large volumes of battery energy storage.
    \item Using hydrogen to provide a significant contribution to the low-carbon heating sector will also require installation of national-scale pipeline infrastructure for both hydrogen (e.g., to supply high-demand areas with production constraints such as London) and for $\mathrm{CO}_2$ (to transport carbon away from hydrogen production towards offshore $\mathrm{CO}_2$ storage sites in the Irish and North Sea).
    \item Reduced usage of hydrogen for heating in 2050-2060 also means that the SMR capacity installed a decade or two earlier gets utilised less. Therefore some of its capacity is used in 2050-2060 to produce hydrogen for running $\mathrm{H}_2$ power generators, which are needed in the power system to ensure sufficient volumes of firm low-carbon generation and provide essential system services such as inertia and frequency regulation.
\end{itemize}

Although the main objective of the paper was to develop a novel method to soft-link two advanced, investment-optimising energy system models, and hence take advantage of their respective strengths, the application of this method on the case of decarbonising the UK’s heat sector also provided several novel insights, as discussed above. Quantitative results point to the cost-effectiveness of using a mix of electricity and hydrogen technologies for delivering zero-carbon heat, also demonstrating a high level of interaction between electricity and hydrogen infrastructure in a zero-carbon system.

Clearly, the quantitative modelling outputs will be affected by the input assumptions made, in particular those regarding the cost of end-use low-carbon heating technologies. A significant volume of sensitivity analysis and further testing will therefore be required in order to provide robust and comprehensive insights into cost-efficient pathways for heat decarbonisation.

One of the key input assumptions for this analysis is the price of natural gas, which has recently seen significant increases. The cost of producing hydrogen using reformer technologies (SMR and ATR) will obviously be highly dependent on the gas price. Higher cost of hydrogen production from gas can be expected to have two main effects: 1)~higher share of electricity in supplying low-carbon heat, and consequently higher capacity of electricity generation and storage, and 2)~improved economics of producing hydrogen from electrolysis compared to reformers, potentially affecting the cost-efficient mix of hydrogen production for the volume that is still a part of the cost-efficient solution.

For a comprehensive assessment of combined electricity and hydrogen supply chains it would be necessary to also consider future hydrogen demand outside of residential and small commercial heating sectors, i.e. in areas such as aviation, shipping, heavy goods transport and industry. Recent projections indicate that this additional demand could reach up to 130~TWh in the UK in 2050 \cite{NGFES2020}. However, these sectors were out of scope of the analysis presented in this paper.

The additional hydrogen demand from industry and transport would also require expanding the capacity for hydrogen production and transport (including the transport of CO$_2$), which might also be required earlier depending on the rate of uptake of hydrogen in these sectors. Sharing the hydrogen and CO$_2$ infrastructure between industry, transport and heating sectors might improve the economics of hydrogen supply.

However, this is not expected to significantly change the preference for hybrid low-carbon heating solutions given the relative costs of supplying electricity and hydrogen and the assumed efficiencies of HP and boiler components, which suggest that HP components should be used to provide the bulk of heat due to low operating cost and high installation cost, while hydrogen boilers should supply peak heat demand because of higher running cost but significantly lower investment cost. Including hydrogen demand from industry and transport sector alongside heating represents an important future extension of the research presented in this paper.

There is a number of other areas where the approach presented in this paper could be further enhanced. Future research in this area will focus on: automating the iteration process between the two models and studying their convergence in more detail; expanding the model scope to include a broader set of technologies for end-use heating and hydrogen production (such as e.g. hydrogen absorption heat pumps), as well as to consider DHNs; and incorporating uncertainty around key variables into the multi-model approach.

\section*{Acknowledgment}
\label{sec:Acknowledgment}
The research presented in this paper has been supported by the UK Engineering and Physical Sciences Research Council (EPSRC) grant number EP/ R045518/1 (IDLES Programme). A shorter version of this paper has been presented during the 16th Conference on Sustainable Development of Energy, Water and Environment Systems (SDEWES) held in Dubrovnik, Croatia, October 10-15, 2021. The title of the conference paper was “Multi-model assessment of heat decarbonisation options in the UK using renewable hydrogen”.

\section*{Appendix}
\label{sec:Appendix}
Input data.

\setcounter{table}{0}
\renewcommand{\thetable}{A\arabic{table}}

\begin{table}[!htbp]
\centering
\caption{Economic parameters of low-carbon and zero-carbon generation technologies}
\resizebox{\textwidth}{!}{%
\begin{tabular}{|l|c|c|c|c|c|}
\hline
\multicolumn{1}{|c|}{\textbf{Generation}} &
  \textbf{\begin{tabular}[c]{@{}c@{}}Capital cost\\  {[}£/kW{]}\end{tabular}} &
  \textbf{\begin{tabular}[c]{@{}c@{}}Fixed O\&M Cost \\ {[}£/kW/year{]}\end{tabular}} &
  \textbf{\begin{tabular}[c]{@{}c@{}}Discount Rate \\ {[}\%{]}\end{tabular}} &
  \textbf{Lifetime} &
  \textbf{\begin{tabular}[c]{@{}c@{}}Carbon Emission \\ {[}kg/MWh{]}\end{tabular}} \\ \hline
Nuclear  & 4100 & 72.9 & 8.9 & 40 & -     \\ \hline
CCGT     &  600 & 13.1 & 7.5 & 25 & 318.8 \\ \hline
OCGT     &  400 &  6.8 & 7.1 & 25 & 520.6 \\ \hline
Gas-CCS  & 1300 & 22.3 & 7.3 & 25 & 31.9  \\ \hline
H2-OCGT  &  400 &  6.8 & 7.1 & 25 & -     \\ \hline
H2-CCGT  &  600 & 13.1 & 7.5 & 25 & -     \\ \hline
Wind     & 1100 & 24.5 & 6.3 & 25 & -     \\ \hline
PV       &  300 &  6.0 & 6.0 & 25 & -     \\ \hline
Battery storage & 395 & - & 7.0 & 20 & -     \\ \hline
\end{tabular}%
}
\end{table}

\begin{table}[!htbp]
\centering
\caption{Annual electricity consumption [TWh]}
\resizebox{\textwidth}{!}{%
\begin{tabular}{|l|ccc|ccc|}
\hline
\multicolumn{1}{|c|}{\multirow{2}{*}{\textbf{Consumer type}}} &
  \multicolumn{3}{c|}{\textbf{$1^\textrm{st}$ iteration}} &
  \multicolumn{3}{c|}{\textbf{$2^\textrm{nd}$ iteration}} \\ \cline{2-7} 
\multicolumn{1}{|c|}{} &
  \multicolumn{1}{c|}{2030-2040} &
  \multicolumn{1}{c|}{2040-2050} &
  2050-2060 &
  \multicolumn{1}{c|}{2030-2040} &
  \multicolumn{1}{c|}{2040-2050} &
  2050-2060 \\ \hline
Non-Heat &
  \multicolumn{1}{c|}{415} &
  \multicolumn{1}{c|}{415} &
  415 &
  \multicolumn{1}{c|}{415} &
  \multicolumn{1}{c|}{415} &
  415 \\ \hline
Heat &
  \multicolumn{1}{c|}{209} &
  \multicolumn{1}{c|}{199} &
  188 &
  \multicolumn{1}{c|}{69} &
  \multicolumn{1}{c|}{97} &
  185 \\ \hline
Total &
  \multicolumn{1}{c|}{624} &
  \multicolumn{1}{c|}{614} &
  603 &
  \multicolumn{1}{c|}{484} &
  \multicolumn{1}{c|}{512} &
  600 \\ \hline
\end{tabular}%
}
\end{table}

\begin{table}[!htbp]
\centering
\caption{Annual electricity production [TWh]}
\resizebox{\textwidth}{!}{%
\begin{tabular}{|l|ccc|ccc|}
\hline
\multicolumn{1}{|c|}{\multirow{2}{*}{\textbf{Producer type}}} &
  \multicolumn{3}{c|}{\textbf{$1^\textrm{st}$ iteration}} &
  \multicolumn{3}{c|}{\textbf{$2^\textrm{nd}$ iteration}} \\ \cline{2-7} 
\multicolumn{1}{|c|}{} &
  \multicolumn{1}{c|}{\textbf{2030-2040}} &
  \multicolumn{1}{c|}{\textbf{2040-2050}} &
  \textbf{2050-2060} &
  \multicolumn{1}{c|}{\textbf{2030-2040}} &
  \multicolumn{1}{c|}{\textbf{2040-2050}} &
  \textbf{2050-2060} \\ \hline
Battery storage $^*$ &
  \multicolumn{1}{c|}{78 - 92 = -14} &
  \multicolumn{1}{c|}{80 - 98 = -18} &
  77 - 95 = -18 &
  \multicolumn{1}{c|}{29 - 34 = -5} &
  \multicolumn{1}{c|}{46 - 56 = -10} &
  74 - 91 = -17 \\ \hline
Other RES &
  \multicolumn{1}{c|}{8} &
  \multicolumn{1}{c|}{8} &
  8 &
  \multicolumn{1}{c|}{8} &
  \multicolumn{1}{c|}{8} &
  8 \\ \hline
H2 generation &
  \multicolumn{1}{c|}{0} &
  \multicolumn{1}{c|}{0} &
  0 &
  \multicolumn{1}{c|}{1} &
  \multicolumn{1}{c|}{21} &
  20 \\ \hline
Biomass &
  \multicolumn{1}{c|}{\textless{}1} &
  \multicolumn{1}{c|}{12} &
  12 &
  \multicolumn{1}{c|}{\textless{}1} &
  \multicolumn{1}{c|}{\textless{}1} &
  \textless{}1 \\ \hline
PV &
  \multicolumn{1}{c|}{26} &
  \multicolumn{1}{c|}{21} &
  20 &
  \multicolumn{1}{c|}{19} &
  \multicolumn{1}{c|}{20} &
  21 \\ \hline
Wind &
  \multicolumn{1}{c|}{494} &
  \multicolumn{1}{c|}{556} &
  545 &
  \multicolumn{1}{c|}{367} &
  \multicolumn{1}{c|}{437} &
  532 \\ \hline
Nuclear &
  \multicolumn{1}{c|}{35} &
  \multicolumn{1}{c|}{35} &
  35 &
  \multicolumn{1}{c|}{35} &
  \multicolumn{1}{c|}{35} &
  35 \\ \hline
OCGT &
  \multicolumn{1}{c|}{3} &
  \multicolumn{1}{c|}{0} &
  0 &
  \multicolumn{1}{c|}{1} &
  \multicolumn{1}{c|}{0} &
  0 \\ \hline
Gas CCGT &
  \multicolumn{1}{c|}{70} &
  \multicolumn{1}{c|}{0} &
  0 &
  \multicolumn{1}{c|}{56} &
  \multicolumn{1}{c|}{0} &
  0 \\ \hline
Total &
  \multicolumn{1}{c|}{624} &
  \multicolumn{1}{c|}{614} &
  603 &
  \multicolumn{1}{c|}{484} &
  \multicolumn{1}{c|}{512} &
  600 \\ \hline
  \multicolumn{7}{l}{\small $^*$ The difference between annual discharged and charged energy for battery storage systems is reported here.}
\end{tabular}%
}
\label{table: Annual_el_production}
\end{table}

\begin{table}[!htbp]
\caption{Cost assumptions for hydrogen and CO$_2$ pipelines of various capacities have been based on \cite{sunny2020needed}, and are listed below.}
\centering
\resizebox{\textwidth}{!}{%
\begin{tabular}{|l|c|c|c|}
\hline
\multicolumn{1}{|c|}{\textbf{Technology type}} & \textbf{CapEx {[}£k/km{]}} & \textbf{Maximum flow rate {[}kg/s{]}} & \textbf{Assumed losses {[}\%/km{]}} \\ \hline
18 inch H$_2$ pipeline  & 870  & 7.1 & \multirow{4}{*}{0.005} \\ \cline{1-3}
24 inch H$_2$ pipeline  & 126  & 30  &                        \\ \cline{1-3}
36 inch H$_2$ pipeline  & 2020 & 105 &                        \\ \cline{1-3}
48 inch H$_2$ pipeline  & 2790 & 220 &                        \\ \hline
12 inch onshore CO$_2$  & 600  & 88  & \multirow{4}{*}{0.002} \\ \cline{1-3}
26 inch onshore CO$_2$  & 1300 & 350 &                        \\ \cline{1-3}
12 inch offshore CO$_2$ & 780  & 88  &                        \\ \cline{1-3}
26 inch offshore CO$_2$                           & \multicolumn{1}{c|}{1500}  & \multicolumn{1}{c|}{350}              &                                     \\ \hline
\end{tabular}%
}
\end{table}

\begin{table}[!htbp]
\caption{The capacity and capital cost for each hydrogen production technology are summarised below. The reduction in capital cost is based on projections by BEIS \cite{BEIS2021H2}. Except for electrolysers, the initial CapEx for 2020-2030 are based on Sunny \textit{et al.} \cite{sunny2020needed}. The CapEx assumptions for electrolysers are based on \cite{CCC2018}.}
\centering
\resizebox{\textwidth}{!}{%
\begin{tabular}{|l|c|cccc|}
\hline
\multicolumn{1}{|c|}{\multirow{2}{*}{\textbf{Technology type}}} &
  \multirow{2}{*}{\textbf{Capacity {[}GW{]}}} &
  \multicolumn{4}{c|}{\textbf{CapEx {[}£/kW{]}}} \\ \cline{3-6} 
\multicolumn{1}{|c|}{} &
   &
  \multicolumn{1}{c|}{\textbf{2020-2030}} &
  \multicolumn{1}{c|}{\textbf{2030-2040}} &
  \multicolumn{1}{c|}{\textbf{2040-2050}} &
  \textbf{2050-2060} \\ \hline
SMR with syngas capture       & 1   & \multicolumn{1}{c|}{320}  & \multicolumn{1}{c|}{280} & \multicolumn{1}{c|}{239} & 199 \\ \hline
SMR with fluegas capture      & 1   & \multicolumn{1}{c|}{480}  & \multicolumn{1}{c|}{420} & \multicolumn{1}{c|}{360} & 299 \\ \hline
ATR with CCS                  & 1   & \multicolumn{1}{c|}{510}  & \multicolumn{1}{c|}{395} & \multicolumn{1}{c|}{331} & 266 \\ \hline
ATR with GHR and CCS          & 1   & \multicolumn{1}{c|}{490}  & \multicolumn{1}{c|}{379} & \multicolumn{1}{c|}{318} & 256 \\ \hline
PEM electrolysis (low eff.)   & 0.1 & \multicolumn{1}{c|}{496}  & \multicolumn{1}{c|}{268} & \multicolumn{1}{c|}{205} & 143 \\ \hline
PEM electrolysis (high eff.)  & 0.1 & \multicolumn{1}{c|}{587}  & \multicolumn{1}{c|}{317} & \multicolumn{1}{c|}{302} & 169 \\ \hline
SOE electrolysis              & 0.1 & \multicolumn{1}{c|}{971}  & \multicolumn{1}{c|}{728} & \multicolumn{1}{c|}{486} & 363 \\ \hline
Biomass gasification with CCS & 0.2 & \multicolumn{1}{c|}{1100} & \multicolumn{1}{c|}{851} & \multicolumn{1}{c|}{713} & 574 \\ \hline
\end{tabular}%
}
\end{table}

\begin{table}[!htbp]
\caption{Assumptions on investment costs (CapEx) and efficiencies of individual heating technologies are reported below based on \cite{andreas_v_olympios_2021_4692649}. The COP refers to an air temperature of 7 $^{\circ}$C and 55 $^{\circ}$C water temperature for an air-water heat pump. Hydrogen boilers are expected to meet the same efficiency standards as current natural gas boilers. The investment cost of the hybrid ASHP-H$_2$ system is constructed from equal shares of the investment cost of ASHP and H$_2$ units.}
\centering
\resizebox{7cm}{!}{%
\begin{tabular}{|l|c|c|}
\hline
\multicolumn{1}{|c|}{\textbf{Technology type}} & \textbf{CapEx {[}£/kWth{]}} & \textbf{\begin{tabular}[c]{@{}c@{}}Efficiency/COP\end{tabular}} \\ \hline
Hydrogen boiler            & 48 & 90\%       \\ \hline
Natural gas boiler & 48 & 90\%       \\ \hline
ASHP              & 412 & 3.04       \\ \hline
Hybrid ASHP-H2             & 209 & 3.03 / 90\% \\ \hline
\end{tabular}%
}
\end{table}

\begin{table}[!htbp]
{\caption{Performance characteristics for hydrogen caverns and CO$_2$ offshore injection wells are summarised here. References underlying these input parameters can be found in \cite{sunny2020needed}. }}
\centering
\resizebox{\textwidth}{!}{%
\begin{tabular}{|l|c|c|c|c|}
\hline
\multicolumn{1}{|c|}{\textbf{Storage type}} & \textbf{CapEx {[}£m/unit{]}} & \textbf{Capacity {[}GWh{]}} & \textbf{Maximum injectivity} & \textbf{Maximum deliverability} \\ \hline
Medium pressure cavern & 32  & 64  & 100         & 200 \\ \hline
High pressure cavern   & 100 & 144 & 100         & 200 \\ \hline
CO$_2$ injection well     & 66  & -   & 1.5 Mt/year & -   \\ \hline
\end{tabular}%
}
\end{table}

\clearpage

\appendix


 \bibliographystyle{elsarticle-num} 
 \bibliography{main}

\end{document}